\documentclass[aps, superscriptaddress, preprint, singlecolumn, longbibliography, showpacs, preprintnumbers,
amssymb, floatfix]{revtex4-2}

\usepackage{graphicx}
\usepackage{epsfig}
\usepackage{amsmath}
\usepackage{amssymb}
\usepackage{amsbsy}
\usepackage{xcolor}
\usepackage{soul}
\usepackage{hyperref}
\usepackage{amsthm}
\usepackage{nicefrac}

\usepackage{setspace}
\usepackage{esint}

\begin{document}

\title{Spatio-temporal characterization of ultrashort vector pulses}
  \date{\today }
  
\author{Apostolos Zdagkas}
\affiliation{Optoelectronics Research Centre and Centre for Photonic Metamaterials, University of Southampton, Southampton SO17 1BJ, United Kingdom}
\author{Venkatram Nalla}
\affiliation{Optoelectronics Research Centre and Centre for Photonic Metamaterials, University of Southampton, Southampton SO17 1BJ, United Kingdom}
\author{Nikitas Papasimakis}
\affiliation{Optoelectronics Research Centre and Centre for Photonic Metamaterials, University of Southampton, Southampton SO17 1BJ, United Kingdom}
\author{Nikolay I. Zheludev}
\affiliation{Optoelectronics Research Centre and Centre for Photonic Metamaterials, University of Southampton, Southampton SO17 1BJ, United Kingdom}
\affiliation{Centre for Disruptive Photonic Technologies, School of Physical and Mathematical Sciences and The Photonics Institute,
            Nanyang Technological University, Singapore 637378, Singapore}

\begin{abstract}
Ultrafast vectorially polarized pulses have found many applications in information and energy transfer owing mainly to the presence of strong longitudinal components and their space-polarization non-separability. Due to their broad spectrum, such pulses often exhibit space-time couplings, which significantly affect the pulse propagation dynamics leading to reduced energy density or utilized to create new effects like a rotating or sliding wavefront at focus. Here, we present a new method for the spatio-temporal characterization of ultrashort cylindrical vector pulses based on a combination of spatially resolved Fourier transform spectroscopy and Mach-Zehnder interferometry. The method provides access to spatially resolved spectral amplitudes and phases of all polarization components of the pulse. We demonstrate the capabilities of the method by completely characterizing a $10$~fs radially polarized pulse from a Ti:sapphire laser at $800$~nm.
\end{abstract}
\maketitle

\section{Introduction}
Space-time couplings (STCs) in propagating waves are defined as the dependence of the temporal properties of the electric field on the transverse spatial coordinates \cite{STCs_review2010}. Mathematically they are revealed as the non-separability of the spatial and temporal terms of the electric field of a pulse into a product, $E(\mathbf{r}, t) \neq f(\mathbf{r})g(t)$. They can significantly affect the energy density of ultrashort pulses at focus \cite{Pretzler2000} since in most cases pulses possessing STCs are not transform limited. STCs also alter the propagation dynamics of ultrashort pulses. The latter has been utilized to create new effects like the lighthouse effect \cite{PhysRevLett.108.113904} where a tilted wavefront is transformed to a rotating wavefront at focus and the ``sliding'' or ``flying'' focus \cite{froula2018spatiotemporal,Sainte-Marie:17} effect where STCs create a focal spot that locally travels with speed greater or lower than the speed of light in free space. 

A range of techniques have been demonstrated for the complete spatio-temporal characterization (retrieval of electric field amplitude and phase) of linearly polarized pulses \cite{Dorrer_review_19}. Such approaches are typically based on scanning the transverse profile of the unknown pulse with a known reference pulse \cite{Bowlan:06,Kosik:05,Wyatt:06} or utilize concepts from wavefront characterization techniques \cite{Cousin:12,Gabolde:06}. Moreover, self-referenced methods have been proposed, where a small part of the pulse under characterization is used as reference. In the latter, the reference is interfered with the unknown pulse and the unknown spectral phase is retrieved through spatially resolved Fourier transform spectroscopy \cite{Miranda:14}. A variation of the latter, named TERMITES \cite{pariente2016space}, uses a slightly expanded replica of the unknown along with an iterative algorithm as a way to reduce the requirements of creating a very homogeneous reference pulse. 

However, none of the aforementioned methods can simultaneously characterize the spatially dependent polarization, intensity and phase that vector polarized pulses exhibit. Indeed, methods such as TERMITES are unsuitable for the characterization of cylindrical vector pulses (CVPs), for example radially polarized pulses, that exhibit polarization singularities at their center. Ultrafast radially polarized pulses have been generated in the femtosecond regime \cite{moh2006direct} and have been compressed down to few femtoseconds \cite{Kong:19, Carbajo:14}. Due to their broad spectrum, such pulses often exhibit space-time couplings (STC). Moreover, space-time couplings have been prescribed to them through a metamaterial converter \cite{PhysRevB.97.201409}. Therefore, their complete characterization becomes of paramount importance for most applications of CVPs. 

\begin{figure}[t]
\centering
\includegraphics[width=\linewidth]{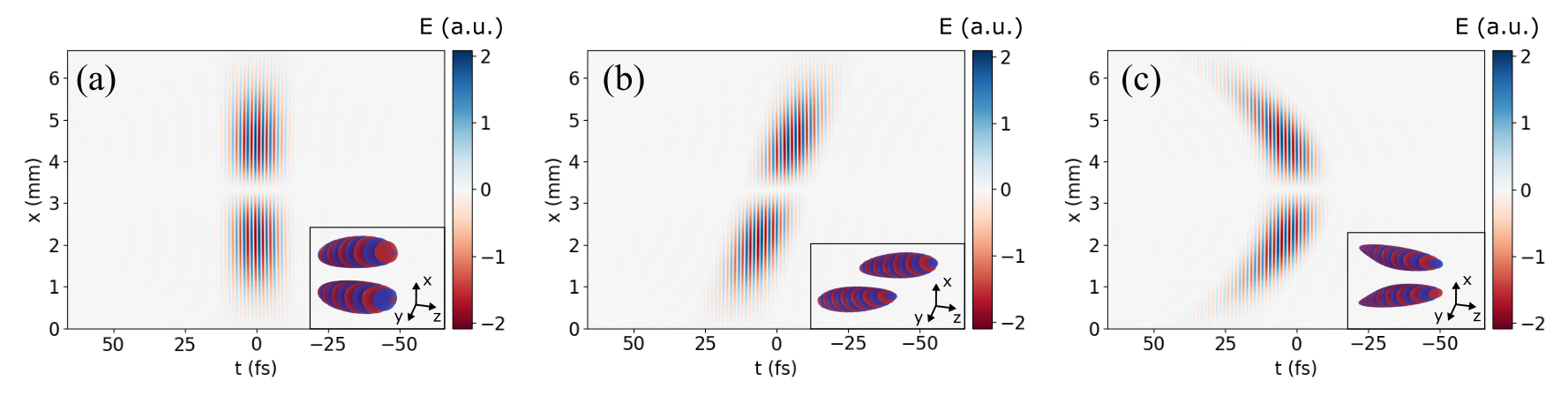}
\caption{\textbf{Characteristic examples of space-time couplings in vector polarized pulses.} Cross sections of the x component of the instantaneous electric field of a radially polarized pulse (a) in the absence of space-time couplings, (b) with pulse front tilt ($3 \; \mathrm{fs/mm}$), and (c) with pulse front curvature ($3 \; \mathrm{fs/mm^2}$). The insets to (a-c) present corresponding 3D illustrations in the form of isosurfaces. The pulse propagates along the positive z axis.}
\label{Fig:stc_radial}
\end{figure}

The characterization of CVPs has been limited mainly to their spatially varying polarization profile. The temporal profile of such pulses has been characterized independently by standard approaches, such as FROG \cite{Trebino:93} and SPIDER \cite{Iaconis:98}. However, such measurements are performed at a single spatial position of the cross section of the pulse, which is far from a complete characterization of the pulses that may exhibit STCs \cite{STCs_review2010}. For example, theoretical calculations have shown that the Flying Donut pulse is isodiffracting \cite{zdagkas2020space}, a property that leads to the spatial profiles of intensity for every frequency component of the pulse to scale along the trajectory of the pulse in the same way. Furthermore, since most ultrafast systems use pulse compressors based on prisms or gratings, in order to implement chirped pulse amplification (CPA) \cite{STRICKLAND1985447}, the appearance of pulse front tilt (PFT) is very likely to occur from small misalignments \cite{STCs_review2010, pft_CPA_lasers} leading to detrimental effects for the pulse duration and its energy density. Such space time couplings occur also in the case of ultrafast CVPs. For example, Fig. \ref{Fig:stc_radial} illustrates the effect of pulse front tilt and pulse front curvature on the shape of a radially polarized pulse. PFT appears as a linearly varying delay of the pulse across a transverse direction while pulse front curvature is a quadratically varying delay from the center of the pulse to its edges. Therefore, the complete characterization of CVPs becomes of paramount importance for most applications.

Only recently a new method has been utilized for the complete spatio-temporal characterization of a 100 fs vector pulse through a two-fold interferometer \cite{alonso2020complete}. Although it was shown to successfully characterize spatially varying polarization gates, it is based on a point scan scheme with an optical fiber, rendering it impractical for the complete transverse spatial characterization of the pulse which would require millions of points to be scanned.

In this work, we present a method for the complete spatio-temporal characterization of CVPs. The method is an extension of the TERMITES technique by means of a Mach-Zehnder interferometer and is termed TERMITES-MAZE (MAch-Zehnder Extended). It allows to fully characterize the spatial and temporal profile of all polarization components of ultrafast CVPs at the spatial resolution of a camera sensor. We illustrate the capabilities of the TERMITES-MAZE technique by applying it to the case of a $10$~fs radially polarized pulse centered at $800$~nm. We show that our approach can reveal the space-time couplings (e.g. pulse front tilt) across different planes for different polarization components of the pulse. We present a detailed description of the experimental implementation of TERMITES-MAZE, while the corresponding algorithm for the analysis of the experimental data is freely available under the BSD 3-clause licence as a python module \cite{azdagkas/TERMITES_MAZE_python_pure}.

As few cycle laser pulses with vectorial fields, orbital angular momentum (OAM) \cite{PhysRevA.45.8185} and other forms of structured light \cite{Rubinsztein_Dunlop_2016} become common, the need for their characterization increases. The presence of STCs become even more important in such broadband pulses that can eventually alter their propagation properties \cite{PhysRevE.65.026606} and their interaction with matter \cite{Hoff_nature,PhysRevLett.124.133202} leading to the appearance of new effects. The tool provided here has the capability to characterize all these types of space-time-polarization ``entanglement'' that are increasingly studied \cite{forbes2019classically} and thus has the potential to accelerate the research and emerging applications based on few cycle vector structured light.

\section{Cylindrical vector pulses}
Cylindrical vector beams are solutions to Maxwell's equation whose amplitude, phase and polarization are axially symmetric. They have been extensively studied and they are routinely generated and characterized \cite{Zhan:09}. Their unique properties have found numerous applications such as tight focusing \cite{PhysRevLett.91.233901}, efficient particle trapping \cite{Michihata:09, Moradi:19}, super-resolution microscopy \cite{Chen:13}, dense optical 3d data storage \cite{Li:11} and data encoding for optical communications \cite{Milione:15} to name a few. Furthermore, their space polarization non-separability has been applied to extend the concepts and tools of quantum physics to ``classically entangled'' states \cite{doi:10.1080/00107514.2019.1580433, mclaren2015measuring}.

Cylindrical vector pulses are the pulsed version of the cylindrical vector beams. Although they offer more opportunities for applications, they are less studied due to the difficulty of their generation and characterization. They share all of the characteristics that make these light structures interesting with additional important features arising due to their pulsed nature. Their short duration renders them ideal for microprocessing of materials since they combine the reduced thermal damage with the increased efficiency and homogeneity of the radial polarization when used in micro-drilling \cite{Allegre_2012}. In fact, single-cycle CVPs, termed "Flying Donuts", exist \cite{PhysRevE.54.889}, for which applications in particle acceleration have been suggested. Moreover, such pulses exhibit a finite topological structure with a finite number of localized singularities \cite{zdagkas2019singularities}, while their toroidal topology is ideal for engaging toroidal and anapole modes in matter \cite{Raybould_interaction,Raybould_anapole_excitation}.

\section{Method}
\label{section:method}
A method for the complete characterization of a CVP should provide all the characteristics of the pulse like the polarization, spectral amplitude and spectral phase at each spatial position. The TERMITES-MAZE approach is capable of retrieving all these characteristics through an interference of a linearly polarized reference pulse with an unknown CVP. Firstly, the TERMITES method is applied for the characterization of the reference pulse and then a Mach-Zehnder interferometer is implemented for the characterization of the unknown. In this way only one polarization of the CVP is characterized at a time. Finally, since all the spatial information is captured, the complete vectorial shape of the pulse can be reconstructed.

\subsection{Experimental implementation}
\label{subsec:experiment}

\begin{figure}[t]
\centering
\includegraphics[width=\linewidth]{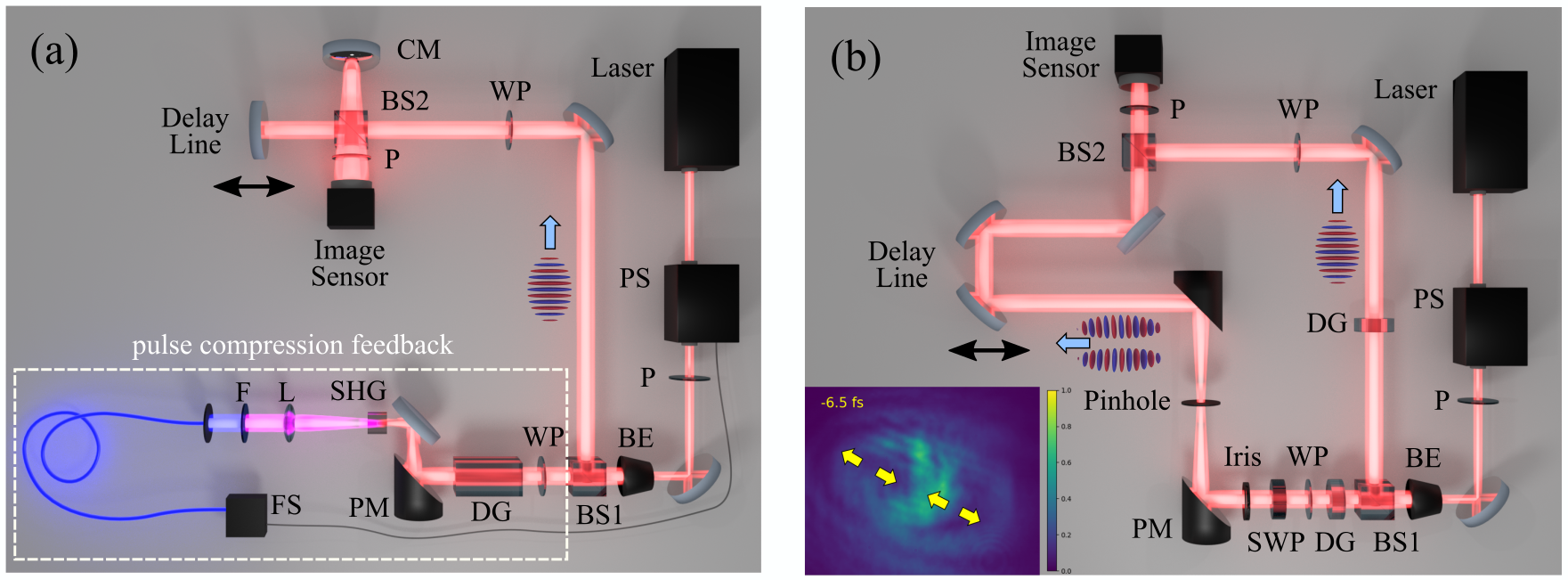}
\caption{Experimental implementation of TERMITES-MAZE. (a) Following the TERMITES technique, a Michelson interferometer setup with a convex mirror in one arm and a delay line in the other is used for the spatio-temporal characterization of a linearly polarized pulse (this pulse will be used as reference). The pulse compression part is shown in the lower part of the setup. (b) A Mach-Zehnder interferometer setup with the one arm carrying the reference pulse (as characterized in (a)) and the other hosting the setup for the generation of the CVP. A delay line is used to scan the unknown pulse with the reference. The generated images are then used to extract the spectral phase and frequency distribution of the unknown pulse. The interference pattern has the expected form of the interference of a linear and a radially polarized pulse. The intensity pattern is antisymmetric with respect to the center of the radially polarized pulse which is a manifestation of the radial polarization shown with yellow arrows. PS: pulse shaper, P: polarizer, BE: beam expander, BS: beam splitter, WP: $\lambda / 2$ waveplate, DG: dispersive glass, PM: parabolic mirror, SHG: second harmonic generation crystal, L: lens, F: filter, FS: fiber spectrometer, CM: convex mirror and SWP: segmented waveplate.}
\label{Fig:setup_tm}
\end{figure}

The TERMITES-MAZE method is an extension of the TERMITES method and as such it involves two experiments. The first experiment is the characterization of a reference linearly polarized pulse using the TERMITES method and the second is a Mach-Zehnder interferometer in which an unknown pulse, in our case a CVP, and the reference pulse are interfered. In both cases, Fig. \ref{Fig:setup_tm}a and b, a $10$ fs Ti:Sa laser with peak frequency at $800$ nm (Spectra-Physics, Element PRO) is used to generate a linearly polarized pulse. The pulse then travels through a pulse shaper that uses a spatial light modulator (SLM) (Biophotonic solutions inc, MIIPS Box640). A polarizer after the pulse shaper is used to select the polarization and to control the intensity of certain frequencies that are set by the SLM. The pulse shaper is used to compress the ultrafast pulse to its Fourier limited duration after travelling though the many optical components that are required for the generation and characterization of the radially polarized pulse. Then a beam expander, doubles the beam width in order to fit the radius of a segmented waveplate that transforms a linear polarized beam to a radial polarized one. After the beam expander, a 50:50 beam splitter (BS1 in Fig.\ref{Fig:setup_tm}) is used to generate two replicas of the pulse. Up to this point the setup is common to both experiments. In the following we describe the two experiments separately. 

The TERMITES experimental setup consists of a simple Michelson interferometer and a digital camera (Thorlabs DCC1545M), see upper part of Fig. \ref{Fig:setup_tm}a. One arm of the interferometer includes a nanometre precision delay line with a scanning range of $20 \mu m$. The other arm includes a convex mirror, whose purpose is to expand the unknown pulse. Hence the image on the camera sensor will be the result of a pulse interfering with an expanding replica of its central part. This expanding central part is considered smoother, in terms of spatio-temporal couplings, than the pulse itself since it is just a small part of the pulse stretched to fit a wider area. A polarizer is finally placed before the camera to increase the signal to noise ratio.

The TERMITES algorithm reveals the spectral phase differences between every point of the pulse and its center. A simple temporal characterization of its central part is the only requirement for a complete spatio-temporal characterization. In our case the ``Multiphoton Intrapulse Interference Phase Scan'' (MIIPS) \cite{Lozovoy:04} technique is used for this purpose since it is already a part of the setup. More specifically, instead of simply characterizing the temporal shape of the pulse, we compress it. The compression will produce a Fourier limited pulse that is also required for the generation of a Fourier limited CVP.

The compression part of the setup can be seen in the lower part of Fig. \ref{Fig:setup_tm}a. The pulse is focused at a second harmonic generation crystal. The generated spectral intensities are then recorded by a spectrometer (Ocean Optics USB2000+XR1-ES) and fed to the pulse shaper's software where the MIIPS method is used for the compression. In the schematic of Fig. \ref{Fig:setup_tm}a, the dispersion compensation glass and the waveplate of the lower arm are used to create the same amount of dispersion with the waveplate of the upper arm, the two passes through the beam splitter and the polarizer. The mirrors have low group delay dispersion and more or less the same number of mirrors is used in the two arms. 

A good spatial resolution to resolve all the fringes and avoid $2\pi$ phase jumps across neighbouring pixels after the Fourier transform is required. Hence a careful choice of the convex mirror induced curvature and the camera sensor is necessary. The same is true for the step of the delay line. Since both spatial and temporal patterns are periodic effects, the experiments are designed to capture more than 10 samples per period, much more than the theoretical minimum according to the sampling theorem \cite{shannon1949communication}.

The above produces a reference (known) linearly polarized pulse. A simple interference of this reference pulse with an unknown more complex structured pulse, like a CVP, is sufficient to reveal the full spatial and temporal structure of the latter. Such an experimental setup is given in Fig. \ref{Fig:setup_tm}b. After the first beam splitter (BS1 in Fig. \ref{Fig:setup_tm}b) two replicas of the pulse are created that travel in two separate arms of a Mach-Zehnder interferometer. The right part of the interferometer carries the reference pulse, while the left arm is used to generate a CVP, which then propagates to the final beam splitter (BS2 in Fig. \ref{Fig:setup_tm}b) where the two pulses recombine and the interference takes place. In this case, the beam splitter has to be polarization independent, as the polarization of the unknown pulse is spatially dependent. As an example, here we characterize a radially polarized pulse that is generated from a segmented waveplate \cite{PhysRevLett.91.233901,Machavariani:07, Carbajo:14}. Our polarization transformer consist of 8 achromatic half-wave plates with the relative orientation of the fast axis at $\pm 11, \pm 34, \pm 56$ and $\pm 79$ degrees. The joints between the waveplates scatter part of the incident light and create diffraction patterns. For that, a $75 \mu m$ pinhole is placed at the Fourier plane to spatially filter the pulse. An iris placed before the focusing mirror is used to select the size of the radial pulse that has to be smaller or equal to the reference. Any diffraction introduced by the iris will be also filtered by the pinhole. A half-waveplate is placed before the segmented waveplate to rotate the input polarization and hence select the output polarization that can be either radial or azimuthal. 

Finally the filtered pulse travels to a delay line that creates a variable delay between the CVP and the reference pulse. The Thorlabs TSGNF5 single axis piezoelectric stage along with a hollow retroreflector were used for the delay line. The interferogram is then captured by the camera. A polarizer before the camera is used not only for the increase of the signal to noise ratio but to characterize the two orthogonal polarizations of the CVP. The latter is achieved by rotating the polarizer at $\pm 45$ degrees with respect to the polarization of the reference pulse. Hence two measurements are required for the complete spatio-temporal characterization of the CVP. The delay between the polarizations is assumed to be zero because of the working principle of the achromatic segmented waveplate. Finally, the dispersion of the pulse must be considered carefully. Both paths need to have the same dispersion as the TERMITES part of the experiment, where the reference pulse was characterized. 

\subsection{Analysis of the recorded data}
\label{subsec:analysis}

\begin{figure}[t]
\centering
\includegraphics[width=\linewidth]{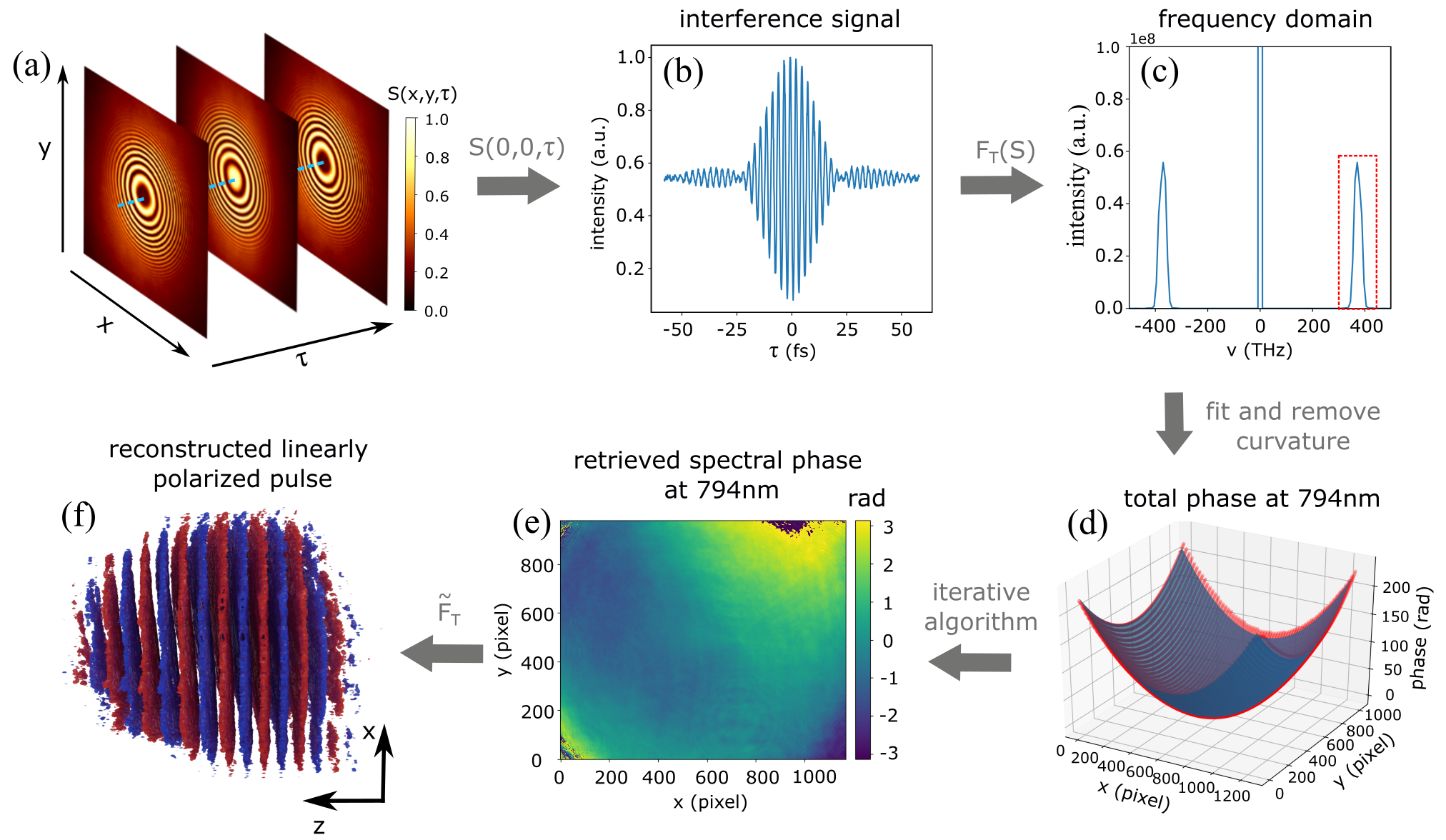}
\caption{A schematic description of the required processing of the images used in the TERMITES algorithm. The data are acquired from the setup described in Fig. \ref{Fig:setup_tm}a. (a) A sequence of interference images captured by a digital camera for successive delays $\tau$ of the reference pulse. (b) The cross-correlation signal, $S(x,y,\tau)$, at a single pixel of the camera sensor. That is the measured intensity as a function of the delay. (c) The spectrum derived by a fast Fourier transform ($F_T$) performed on the cross-correlation signal of the pixel. (d) Unwrapped phase data (red dots) and the quadratic fitted surface (blue) close to the central wavelength ($\simeq 800$ nm). The fitted phase is removed from the total spatially depended phase to reveal the phase due to the STC. (e) The retrieved phase at $794$ nm after the application of the iterative algorithm. (f) The 3d reconstruction of the pulse in time and space through an inverse Fourier transform ($\tilde{F}_T$) of the retrieved amplitude and phase data.}
\label{Fig:algorithm}
\end{figure}

In this section we briefly discuss the TERMITES algorithm and we present the extra analysis steps needed  for the characterization of CVPs. The TERMITES method is used for the spatiotemporal characterization of linearly polarized pulses up to a constant spectral phase and an ambiguity in the pulse front curvature. Our approach inherits these limitations.

\begin{figure}[t]
\centering
\includegraphics[width=\linewidth]{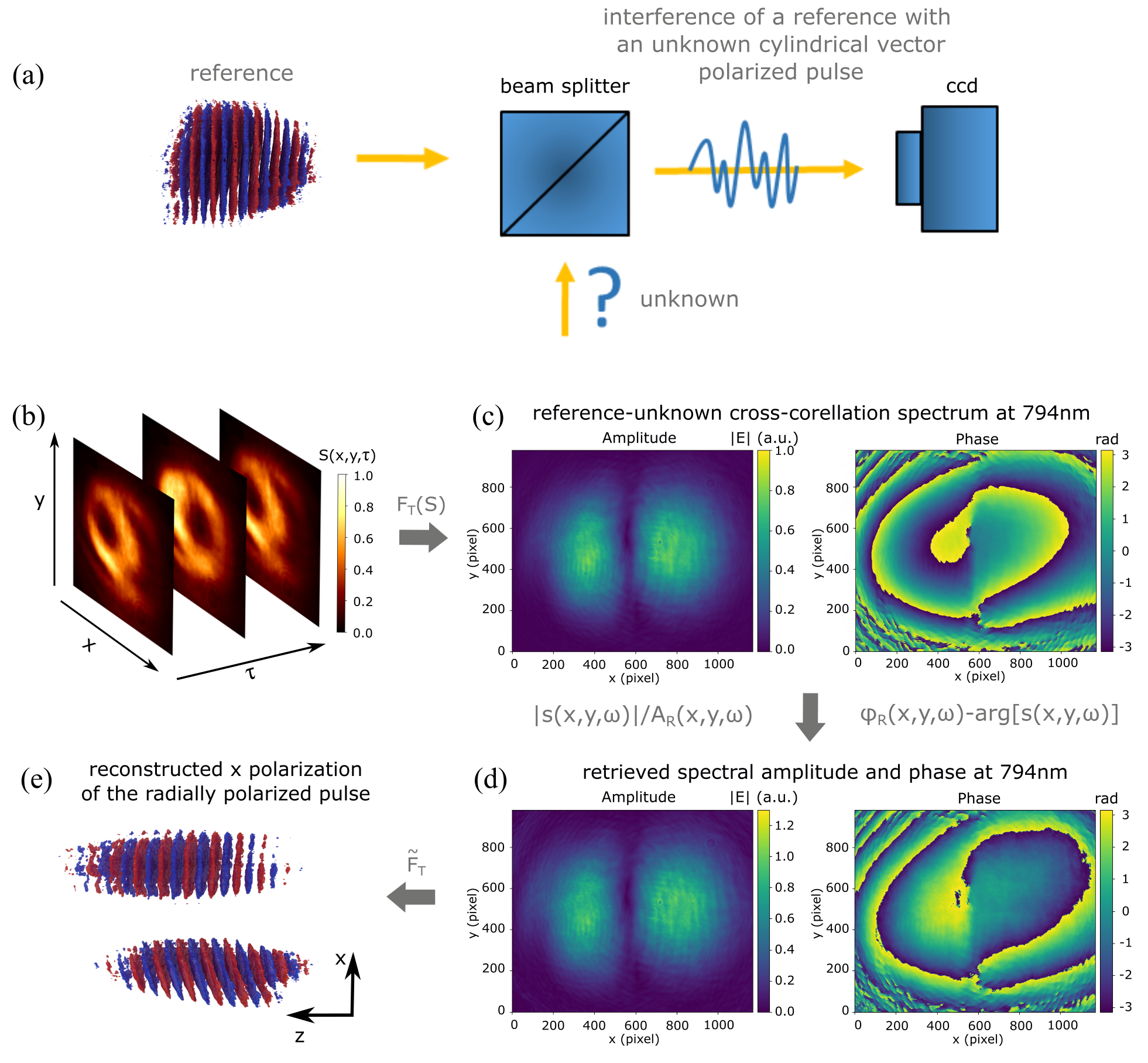}
\caption{The final step of TERMITES-MAZE method for the characterization of a radially polarized pulse. (a) A simplified schematic of the experiment where the reference pulse from the previous part of the method is interfered with the unknown CVP. (b) A sequence of interference images captured by a digital camera for successive delays of the reference pulse.(c) The Fourier transformed cross-correlation signal close to the central wavelength of the pulse. The retrieved amplitude and phase of the x polarization of the radially polarized pulse at $794$ nm after the division and subtraction of the reference pulse's amplitude and phase from the cross-correlation spectra. (d) The 3d reconstruction of the x polarization of the radially polarized pulse in time and space.}
\label{Fig:termites_maze}
\end{figure}

In TERMITES, a pulse is interfered with a magnified replica of itself. A delay line is used to scan the unknown pulse with the reference. The interference pattern is then recorded by a digital camera and an image for each relative delay is stored in a computer, resulting in a 3D dataset (see Fig. \ref{Fig:algorithm}a). For each pixel across all images an interferogram is captured which is the collected energy during the exposure period transferred to the sensor by the two pulses, as a function of the delay. This signal is proportional to the time integral of the intensity of the total electric field and can be written as
\begin{align}
S(\mathbf{r},\tau) &= \int |E_{\mathrm{R}}(\mathbf{r},t) + E(\mathbf{r},t-\tau)|^2 dt \nonumber \\ 
&= I(\mathbf{r}) + I_{\mathrm{R}}(\mathbf{r}) + \int [E_{\mathrm{R}}(\mathbf{r},t)E^*(\mathbf{r},t-\tau) + E^*_{\mathrm{R}}(\mathbf{r},t)E(\mathbf{r},t-\tau)] dt,
\label{Eq:cross_cor}
\end{align}
with the subscript ``$\mathrm{R}$'' denoting the reference (diverging) pulse and $\tau$ the delay. The signal on a central pixel is shown in Fig. \ref{Fig:algorithm}b. A constant signal is observed when there is no overlap of the pulses, mathematically described by the first two terms of Eq. \ref{Eq:cross_cor}, and an interference signal when they overlap, mathematically described by the integral. The term $s(\mathbf{r},\tau) = \int E_{\mathrm{R}}(\mathbf{r},t)E^*(\mathbf{r},t-\tau)dt$ is the cross-correlation of the two fields. Hence Eq. \ref{Eq:cross_cor} takes the form
\begin{align}
S(\mathbf{r},\tau)= I(\mathbf{r}) + I_{\mathrm{R}}(\mathbf{r}) + s(\mathbf{r},\tau) + s^*(\mathbf{r},\tau).
\label{Eq:cross_cor2}
\end{align}

\begin{figure}[t]
\centering
\includegraphics[width=\linewidth]{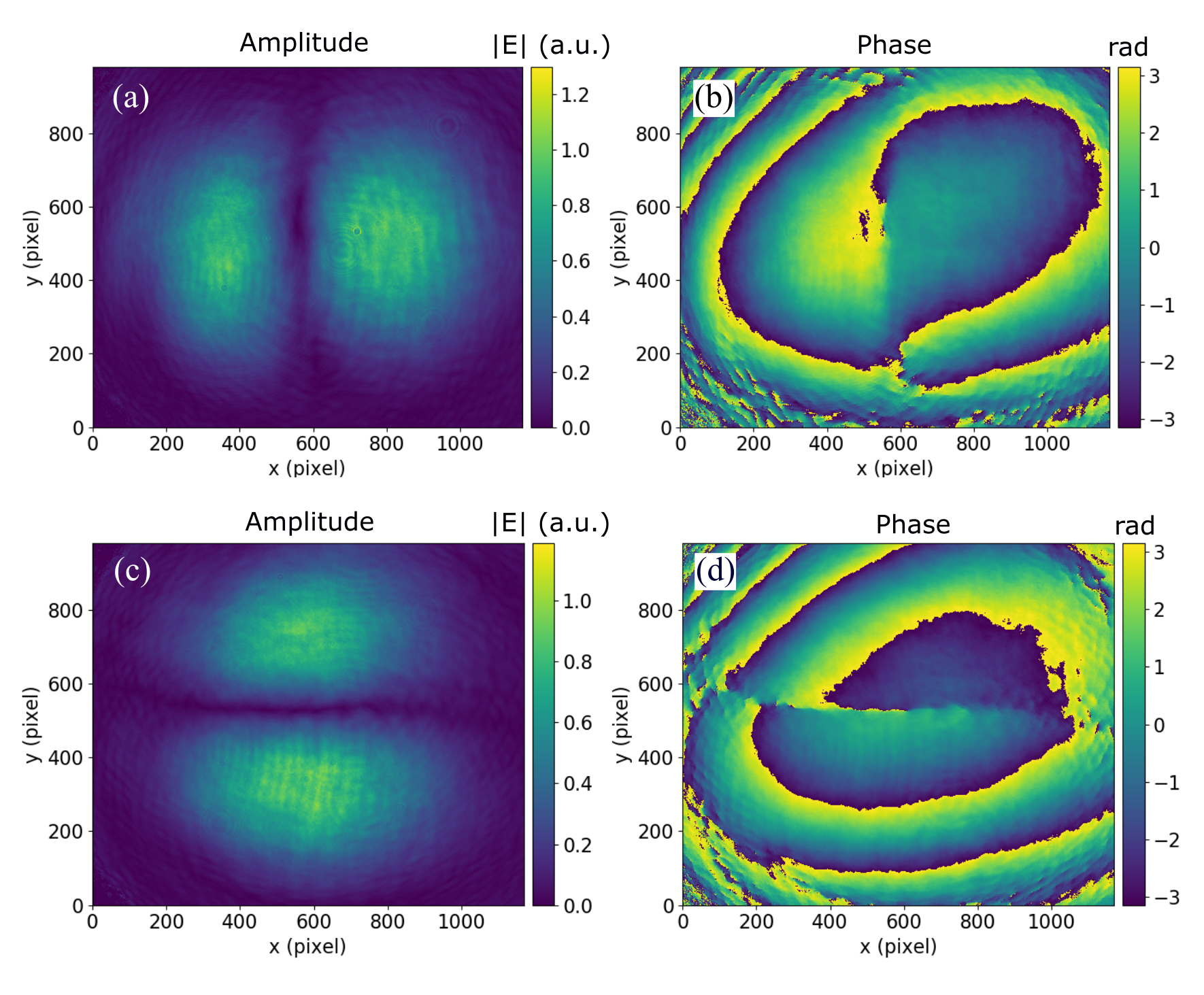}
\caption{(a) and (c) Retrieved amplitude of the horizontally and vertically polarized areas of a radially polarized pulse at $794$ nm respectively. (b) and (d) Retrieved phase at the same wavelength for the horizontal (vertical) polarization according to the TERMITES-MAZE algorithm. The spectral phase has a difference of $\pi$ between the left and right (top and bottom) areas as it is expected for a radially polarized pulse.}
\label{Fig:amp_phase_mz}
\end{figure}

A Fourier transform is then performed for every pixel and the frequency domain signal is now described by the equation
\begin{align}
S(\mathbf{r}, \omega)= F_{\mathrm{T}}[S(\mathbf{r}, \tau)] = F_{\mathrm{T}}[I(\mathbf{r}) + I_{\mathrm{R}}(\mathbf{r})] + s(\mathbf{r},\omega) + s^*(\mathbf{r},-\omega),
\label{Eq:cross_cor_fd}
\end{align}
with $F_{\mathrm{T}}$ denoting the time to frequency Fourier transform operator. The equation describes the presence of a zero frequency term, derived from the Fourier transform of the time stationary term, and two frequency signals with equal amplitude and at symmetric frequencies. Fig. \ref{Fig:algorithm}c shows the results derived from the above analysis regarding a central pixel of the camera which corresponds to the central part of the pulse. A zero frequency term and two symmetric peaks with central wavelength at around $800$ nm, in accordance with the central wavelength of our laser, are formed by the Fourier transform.

The result of the above procedure leads to two sets of images. One carries the intensity distribution at each frequency and the other the spectral phase. The negative frequencies can be derived by the complex conjugate of the positive and hence are redundant for the analysis. Additionally, the noise free spectrum is from about 700 to 900 nm. Only a few images are thus needed for the analysis and hence the processing time is reduced significantly without any loss of information.

The Fourier transform of the cross-correlation of two signals, $E(\mathbf{r},t)$ and $E_{\mathrm{R}}(\mathbf{r},t)$, is equal to the product of their spectra with one of them being complex conjugate, as in the cross-correlation integrand. In our case we have $s(\mathbf{r},\omega) = E_{\mathrm{R}}(\mathbf{r},\omega) E^*(\mathbf{r},\omega)$. The amplitude and phase of the signal that we are interested are thus given by
\begin{align}
|s(\mathbf{r},\omega)| &= A_{\mathrm{R}}(\mathbf{r},\omega)A(\mathbf{r},\omega)
\label{Eq:cross_cor_amplitude} \\
\mathrm{arg}(s(\mathbf{r},\omega))&= \varphi_{\mathrm{R}}  (\mathbf{r},\omega) - \varphi(\mathbf{r},\omega)
\label{Eq:cross_cor_arg}
\end{align}
with $A$ being the amplitude of the spectrum and $\varphi$ the spectral phase.

In the TERMITES algorithm, the spectral phase that is created from the convex mirror is removed by fitting a quadratic surface to the phase data, Fig. \ref{Fig:algorithm}d and then an iterative algorithm is applied Fig. \ref{Fig:algorithm}e until the retrieved spectral phase can correctly reproduce the known relationship between the pulse and its expanding replica \cite{pariente2016space}. An inverse Fourier transform then reconstructs the linearly polarized pulse in time and space, Fig. \ref{Fig:algorithm}f. Following the characterization of the reference pulse the Mach-Zehnder experiment is performed. The analysis is the same up to Eq. \ref{Eq:cross_cor_amplitude} and \ref{Eq:cross_cor_arg}. However, in this case the reference spectral amplitude and spectral phase are known, hence a simple division of the spectral amplitudes and a subtraction of the spectral phases gives the complete frequency domain representation of the unknown CVP. The procedure is then repeated for the orthogonal polarization and the results are combined.

\begin{figure}[t]
\centering
\includegraphics[width=0.9\linewidth]{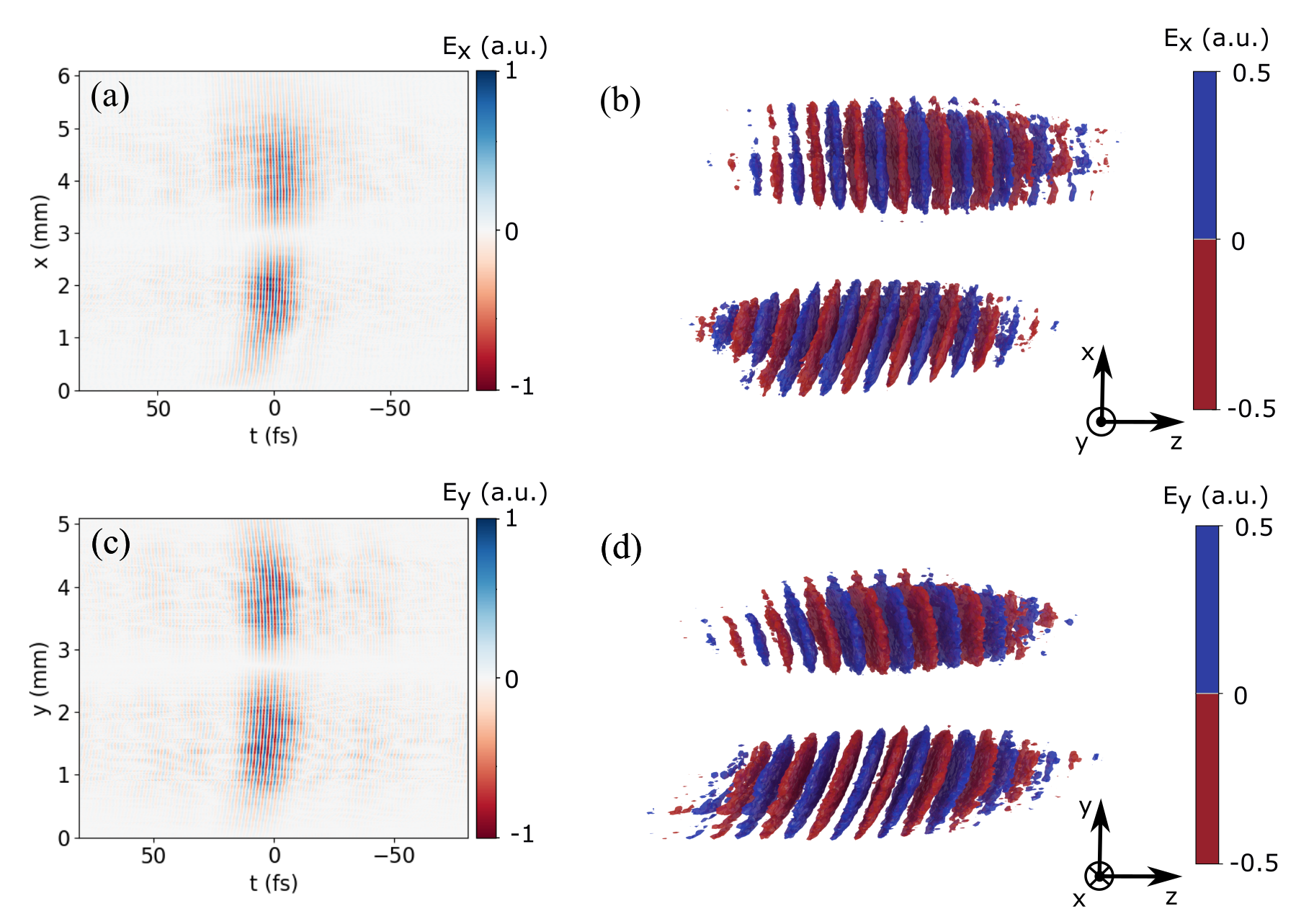}
\caption{(a)-(b) $x-t$ cross-section of the x polarized electric field of a 10 fs radially polarized pulse and its isosurface plot at half of the maximum electric field value as it is reconstructed with the TERMITES-MAZE method. (c)-(d) $y-t$ cross-section of the y polarized electric field of the same pulse and its isosurface plot.}
\label{Fig:reconstruct_radial}
\end{figure}

\begin{figure}[t]
\centering
\includegraphics[width=0.9\linewidth]{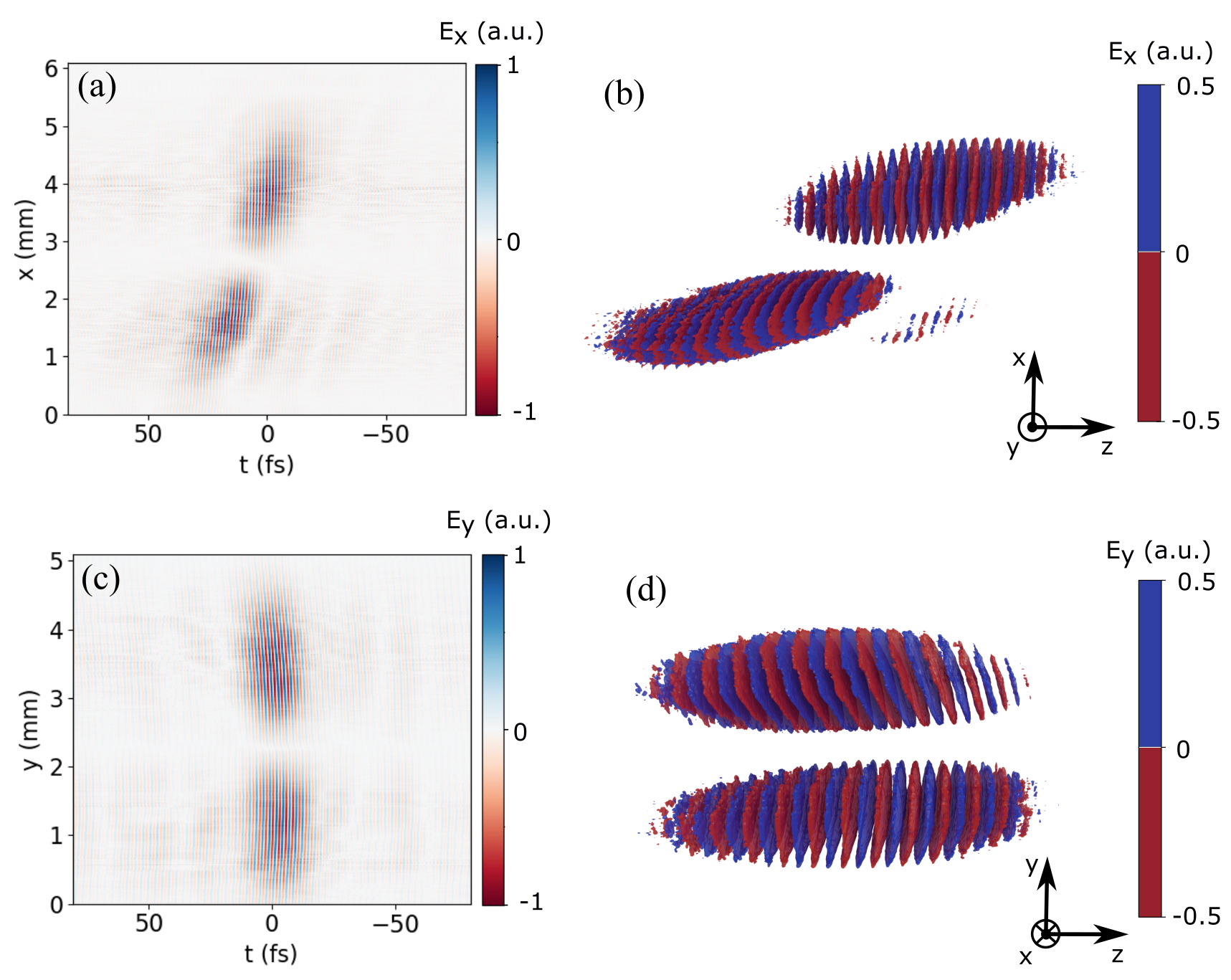}
\caption{(a-b) $x-t$ cross-section of the x polarized electric field of a 10 fs radially polarized pulse exhibiting PFT and its isosurface plot at half of the maximum electric field value as it is reconstructed with the TERMITES-MAZE method. (c-d) $y-t$ cross-section of the y polarized electric field of the same pulse and its isosurface plot.}
\label{Fig:reconstruct_radial_pft}
\end{figure}

Due to the nature of the experiments, the pulses must be very well aligned. This is particularly important for the alignment between the two experiments. In order to achieve high precision, we perform the alignment with image processing. A fit of a two-dimensional elliptical Gaussian function to the intensity of the same pulse as captured by the sensor at the two different experiments, gives the position of its center, the FWHM at different orientations and the rotation angle. Since the pulse is the same, a comparison of these values provides the exact spatial shift of the pulse on the sensor. Then by cropping and flipping the corresponding images the two experiments are aligned with pixel size precision.  

Fig. \ref{Fig:amp_phase_mz} shows the spectral amplitude and spectral phase of the $x$ and $y$ polarizations of a radially polarized pulse at $794$ nm as it is derived by the analysis. Only areas with significant energy are considered for the phase maps. In these areas the pulse has a phase difference of $\pi$ between its left and right (or top-bottom) parts as it is expected for a CVP due to the polarization singularity in its center. The small deviation of the phase profile from the cylindrical symmetry is due to the existence of relatively weak space-time couplings, like pulse front tilt and curvature as it can be seen in Fig. \ref{Fig:reconstruct_radial}.

An inverse Fourier transform is used to reconstruct the 3 dimensional shape of the pulse in the time domain. Fig \ref{Fig:reconstruct_radial}a-d show the $x-t$ and $y-t$ cross-sections for the $x$ and $y$ polarization respectively of a radially polarized pulse and their 3d reconstruction as an isosurface at half of the maximum amplitude. The pulse is 10 fs long and it is clear that apart from a small curvature and a small pulse front tilt in the $x$ direction, it does not have significant aberrations. 

In contrast, in Fig. \ref{Fig:reconstruct_radial_pft} we show the reconstruction of a pulse exhibiting PFT in its $x$ direction. This is actually the initial pulse that was used for the experiment and the PFT indicates a misalignment of the pulse compression setup. The existence of the PFT offers an opportunity to apply the developed method for the characterization and correction of the latter. The measured PFT is found to be about 7 $\mathrm{fs/mm}$ and it was retrieved by fitting a paraboloid to the spatially depended delay of the pulse which is directly related to the spatially depended spectral phase provided by the analysis \cite{jolly2020spatio}. A way to quickly correct the tilt is to insert a wedge. The induced PFT due to a wedge can be calculated from the following equation \cite{Dorrer_review_19}
\begin{align}
\xi = \frac{\lambda_0 \tan (\theta)}{c} \frac{\mathrm{d} n}{\mathrm{d} \lambda}
\label{Eq:pft_wedge} 
\end{align}
with $\lambda_0$ the central wavelength of the pulse, $c$ the speed of light, $n$ the refractive index and $\theta$ the angle of the wedge. Here, we have used a commercially available $5^{\circ}$ fused silica wedge plate in the path of the pulse. From Eq. \ref{Eq:pft_wedge} we calculate the PFT introduced by the wedge to be about 6 $\mathrm{fs/mm}$. The wedge was inserted in such a way that influenced and corrected the profile of the pulse along the horizontal axis, $x$ direction. The corrected pulse is shown in the Fig. \ref{Fig:reconstruct_radial} and exhibits a PFT of about 0.5 $\mathrm{fs/mm}$, which is close to the expected value from Eq. \ref{Eq:pft_wedge}. The fact that the existence of the PFT on the generated CVP was discovered only during the development of the method highlights its practicality to any field employing such pulses.

\section{Conclusions}
In this paper we have presented a technique for the spatio-temporal characterization of CVPs and we demonstrated its capabilities with the complete spatio-temporal characterization of a 10 fs radially polarized pulse. The technique is a Mach-Zehnder interferometer extension to the TERMITES method that is necessary to overcome the limitation of the latter to characterize pulses with polarization singularities. A radially polarized pulse exhibiting pulse front tilt was also characterized to illustrate the capabilities of the method.

In the current implementation, the method characterizes CVPs whose relative phase between the two orthogonal polarizations is a priory known and hence it is not measured. However, there are cases that this relative phase is not known and it is the important quantity. Such cases are pulses with dynamic polarization like polarization gates that are used for the generation of isolated attosecond pulses \cite{sola2006controlling} or pulses with polarization singularities that are unknown to exist a priori. The TERMITES-MAZE method can be easily extended for the characterization of such pulses by altering the setup to use a reference pulse polarized at 45 degrees and interfere it with the x and y components of the unknown pulse on a polarizing beam splitter, similar to the POLLIWOG method \cite{Walecki:97}.

The method is not limited by the bandwidth of the pulse since it is based on an interferometric setup. Apart from the characterization of CVPs it can be applied to ultrashort light structures possessing a number of polarization or phase singularities. The latter is of paramount importance to transfer the field of topological optics \cite{dennis_chapter_2009} from monochromatic beams to singular pulses \cite{zdagkas2019singularities}. Additionally, all applications of CVPs such as free space communications or micro-machining that depend on the propagation dynamics or light matter interactions can directly benefit from the method acting as a feedback that allows to tune the pulses to have the optimal shape or exploring the generation and effects of new light structures to already existing applications.

Finally, a software package for the analysis of the experimental data has been developed in the python programming language and is freely available under the BSD 3-clause licence \cite{azdagkas/TERMITES_MAZE_python_pure}.

\bibliography{termites_maze_bib}

\begin{thebibliography}{51}%
\makeatletter
\providecommand \@ifxundefined [1]{%
 \@ifx{#1\undefined}
}%
\providecommand \@ifnum [1]{%
 \ifnum #1\expandafter \@firstoftwo
 \else \expandafter \@secondoftwo
 \fi
}%
\providecommand \@ifx [1]{%
 \ifx #1\expandafter \@firstoftwo
 \else \expandafter \@secondoftwo
 \fi
}%
\providecommand \natexlab [1]{#1}%
\providecommand \enquote  [1]{``#1''}%
\providecommand \bibnamefont  [1]{#1}%
\providecommand \bibfnamefont [1]{#1}%
\providecommand \citenamefont [1]{#1}%
\providecommand \href@noop [0]{\@secondoftwo}%
\providecommand \href [0]{\begingroup \@sanitize@url \@href}%
\providecommand \@href[1]{\@@startlink{#1}\@@href}%
\providecommand \@@href[1]{\endgroup#1\@@endlink}%
\providecommand \@sanitize@url [0]{\catcode `\\12\catcode `\$12\catcode
  `\&12\catcode `\#12\catcode `\^12\catcode `\_12\catcode `\%12\relax}%
\providecommand \@@startlink[1]{}%
\providecommand \@@endlink[0]{}%
\providecommand \url  [0]{\begingroup\@sanitize@url \@url }%
\providecommand \@url [1]{\endgroup\@href {#1}{\urlprefix }}%
\providecommand \urlprefix  [0]{URL }%
\providecommand \Eprint [0]{\href }%
\providecommand \doibase [0]{https://doi.org/}%
\providecommand \selectlanguage [0]{\@gobble}%
\providecommand \bibinfo  [0]{\@secondoftwo}%
\providecommand \bibfield  [0]{\@secondoftwo}%
\providecommand \translation [1]{[#1]}%
\providecommand \BibitemOpen [0]{}%
\providecommand \bibitemStop [0]{}%
\providecommand \bibitemNoStop [0]{.\EOS\space}%
\providecommand \EOS [0]{\spacefactor3000\relax}%
\providecommand \BibitemShut  [1]{\csname bibitem#1\endcsname}%
\let\auto@bib@innerbib\@empty
\bibitem [{\citenamefont {Akturk}\ \emph {et~al.}(2010)\citenamefont {Akturk},
  \citenamefont {Gu}, \citenamefont {Bowlan},\ and\ \citenamefont
  {Trebino}}]{STCs_review2010}%
  \BibitemOpen
  \bibfield  {author} {\bibinfo {author} {\bibfnamefont {S.}~\bibnamefont
  {Akturk}}, \bibinfo {author} {\bibfnamefont {X.}~\bibnamefont {Gu}}, \bibinfo
  {author} {\bibfnamefont {P.}~\bibnamefont {Bowlan}},\ and\ \bibinfo {author}
  {\bibfnamefont {R.}~\bibnamefont {Trebino}},\ }\bibfield  {title} {\bibinfo
  {title} {Spatio-temporal couplings in ultrashort laser pulses},\ }\href
  {http://stacks.iop.org/2040-8986/12/i=9/a=093001} {\bibfield  {journal}
  {\bibinfo  {journal} {Journal of Optics}\ }\textbf {\bibinfo {volume} {12}},\
  \bibinfo {pages} {093001} (\bibinfo {year} {2010})}\BibitemShut {NoStop}%
\bibitem [{\citenamefont {Pretzler}\ \emph
  {et~al.}(2000{\natexlab{a}})\citenamefont {Pretzler}, \citenamefont
  {Kasper},\ and\ \citenamefont {Witte}}]{Pretzler2000}%
  \BibitemOpen
  \bibfield  {author} {\bibinfo {author} {\bibfnamefont {G.}~\bibnamefont
  {Pretzler}}, \bibinfo {author} {\bibfnamefont {A.}~\bibnamefont {Kasper}},\
  and\ \bibinfo {author} {\bibfnamefont {K.}~\bibnamefont {Witte}},\ }\bibfield
   {title} {\bibinfo {title} {Angular chirp and tilted light pulses in cpa
  lasers},\ }\href {https://doi.org/10.1007/s003400050001} {\bibfield
  {journal} {\bibinfo  {journal} {Applied Physics B: Lasers and Optics}\
  }\textbf {\bibinfo {volume} {70}},\ \bibinfo {pages} {1} (\bibinfo {year}
  {2000}{\natexlab{a}})}\BibitemShut {NoStop}%
\bibitem [{\citenamefont {Vincenti}\ and\ \citenamefont
  {Qu\'er\'e}(2012)}]{PhysRevLett.108.113904}%
  \BibitemOpen
  \bibfield  {author} {\bibinfo {author} {\bibfnamefont {H.}~\bibnamefont
  {Vincenti}}\ and\ \bibinfo {author} {\bibfnamefont {F.}~\bibnamefont
  {Qu\'er\'e}},\ }\bibfield  {title} {\bibinfo {title} {Attosecond lighthouses:
  How to use spatiotemporally coupled light fields to generate isolated
  attosecond pulses},\ }\href {https://doi.org/10.1103/PhysRevLett.108.113904}
  {\bibfield  {journal} {\bibinfo  {journal} {Phys. Rev. Lett.}\ }\textbf
  {\bibinfo {volume} {108}},\ \bibinfo {pages} {113904} (\bibinfo {year}
  {2012})}\BibitemShut {NoStop}%
\bibitem [{\citenamefont {Froula}\ \emph {et~al.}(2018)\citenamefont {Froula},
  \citenamefont {Turnbull}, \citenamefont {Davies}, \citenamefont {Kessler},
  \citenamefont {Haberberger}, \citenamefont {Palastro}, \citenamefont {Bahk},
  \citenamefont {Begishev}, \citenamefont {Boni}, \citenamefont {Bucht} \emph
  {et~al.}}]{froula2018spatiotemporal}%
  \BibitemOpen
  \bibfield  {author} {\bibinfo {author} {\bibfnamefont {D.~H.}\ \bibnamefont
  {Froula}}, \bibinfo {author} {\bibfnamefont {D.}~\bibnamefont {Turnbull}},
  \bibinfo {author} {\bibfnamefont {A.~S.}\ \bibnamefont {Davies}}, \bibinfo
  {author} {\bibfnamefont {T.~J.}\ \bibnamefont {Kessler}}, \bibinfo {author}
  {\bibfnamefont {D.}~\bibnamefont {Haberberger}}, \bibinfo {author}
  {\bibfnamefont {J.~P.}\ \bibnamefont {Palastro}}, \bibinfo {author}
  {\bibfnamefont {S.-W.}\ \bibnamefont {Bahk}}, \bibinfo {author}
  {\bibfnamefont {I.~A.}\ \bibnamefont {Begishev}}, \bibinfo {author}
  {\bibfnamefont {R.}~\bibnamefont {Boni}}, \bibinfo {author} {\bibfnamefont
  {S.}~\bibnamefont {Bucht}}, \emph {et~al.},\ }\bibfield  {title} {\bibinfo
  {title} {Spatiotemporal control of laser intensity},\ }\href@noop {}
  {\bibfield  {journal} {\bibinfo  {journal} {Nature Photonics}\ }\textbf
  {\bibinfo {volume} {12}},\ \bibinfo {pages} {262} (\bibinfo {year}
  {2018})}\BibitemShut {NoStop}%
\bibitem [{\citenamefont {Sainte-Marie}\ \emph {et~al.}(2017)\citenamefont
  {Sainte-Marie}, \citenamefont {Gobert},\ and\ \citenamefont
  {Quere}}]{Sainte-Marie:17}%
  \BibitemOpen
  \bibfield  {author} {\bibinfo {author} {\bibfnamefont {A.}~\bibnamefont
  {Sainte-Marie}}, \bibinfo {author} {\bibfnamefont {O.}~\bibnamefont
  {Gobert}},\ and\ \bibinfo {author} {\bibfnamefont {F.}~\bibnamefont
  {Quere}},\ }\bibfield  {title} {\bibinfo {title} {Controlling the velocity of
  ultrashort light pulses in vacuum through spatio-temporal couplings},\ }\href
  {https://doi.org/10.1364/OPTICA.4.001298} {\bibfield  {journal} {\bibinfo
  {journal} {Optica}\ }\textbf {\bibinfo {volume} {4}},\ \bibinfo {pages}
  {1298} (\bibinfo {year} {2017})}\BibitemShut {NoStop}%
\bibitem [{\citenamefont {{Dorrer}}(2019)}]{Dorrer_review_19}%
  \BibitemOpen
  \bibfield  {author} {\bibinfo {author} {\bibfnamefont {C.}~\bibnamefont
  {{Dorrer}}},\ }\bibfield  {title} {\bibinfo {title} {Spatiotemporal metrology
  of broadband optical pulses},\ }\href
  {https://doi.org/10.1109/JSTQE.2019.2899019} {\bibfield  {journal} {\bibinfo
  {journal} {IEEE Journal of Selected Topics in Quantum Electronics}\ }\textbf
  {\bibinfo {volume} {25}},\ \bibinfo {pages} {1} (\bibinfo {year}
  {2019})}\BibitemShut {NoStop}%
\bibitem [{\citenamefont {Bowlan}\ \emph {et~al.}(2006)\citenamefont {Bowlan},
  \citenamefont {Gabolde}, \citenamefont {Shreenath}, \citenamefont
  {McGresham}, \citenamefont {Trebino},\ and\ \citenamefont
  {Akturk}}]{Bowlan:06}%
  \BibitemOpen
  \bibfield  {author} {\bibinfo {author} {\bibfnamefont {P.}~\bibnamefont
  {Bowlan}}, \bibinfo {author} {\bibfnamefont {P.}~\bibnamefont {Gabolde}},
  \bibinfo {author} {\bibfnamefont {A.}~\bibnamefont {Shreenath}}, \bibinfo
  {author} {\bibfnamefont {K.}~\bibnamefont {McGresham}}, \bibinfo {author}
  {\bibfnamefont {R.}~\bibnamefont {Trebino}},\ and\ \bibinfo {author}
  {\bibfnamefont {S.}~\bibnamefont {Akturk}},\ }\bibfield  {title} {\bibinfo
  {title} {Crossed-beam spectral interferometry: a simple,
  high-spectral-resolution method for completely characterizing complex
  ultrashort pulses in real time},\ }\href
  {https://doi.org/10.1364/OE.14.011892} {\bibfield  {journal} {\bibinfo
  {journal} {Opt. Express}\ }\textbf {\bibinfo {volume} {14}},\ \bibinfo
  {pages} {11892} (\bibinfo {year} {2006})}\BibitemShut {NoStop}%
\bibitem [{\citenamefont {Kosik}\ \emph {et~al.}(2005)\citenamefont {Kosik},
  \citenamefont {Radunsky}, \citenamefont {Walmsley},\ and\ \citenamefont
  {Dorrer}}]{Kosik:05}%
  \BibitemOpen
  \bibfield  {author} {\bibinfo {author} {\bibfnamefont {E.~M.}\ \bibnamefont
  {Kosik}}, \bibinfo {author} {\bibfnamefont {A.~S.}\ \bibnamefont {Radunsky}},
  \bibinfo {author} {\bibfnamefont {I.~A.}\ \bibnamefont {Walmsley}},\ and\
  \bibinfo {author} {\bibfnamefont {C.}~\bibnamefont {Dorrer}},\ }\bibfield
  {title} {\bibinfo {title} {Interferometric technique for measuring broadband
  ultrashort pulses at the sampling limit},\ }\href
  {https://doi.org/10.1364/OL.30.000326} {\bibfield  {journal} {\bibinfo
  {journal} {Opt. Lett.}\ }\textbf {\bibinfo {volume} {30}},\ \bibinfo {pages}
  {326} (\bibinfo {year} {2005})}\BibitemShut {NoStop}%
\bibitem [{\citenamefont {Wyatt}\ \emph {et~al.}(2006)\citenamefont {Wyatt},
  \citenamefont {Walmsley}, \citenamefont {Stibenz},\ and\ \citenamefont
  {Steinmeyer}}]{Wyatt:06}%
  \BibitemOpen
  \bibfield  {author} {\bibinfo {author} {\bibfnamefont {A.~S.}\ \bibnamefont
  {Wyatt}}, \bibinfo {author} {\bibfnamefont {I.~A.}\ \bibnamefont {Walmsley}},
  \bibinfo {author} {\bibfnamefont {G.}~\bibnamefont {Stibenz}},\ and\ \bibinfo
  {author} {\bibfnamefont {G.}~\bibnamefont {Steinmeyer}},\ }\bibfield  {title}
  {\bibinfo {title} {Sub-10 fs pulse characterization using spatially encoded
  arrangement for spectral phase interferometry for direct electric field
  reconstruction},\ }\href {https://doi.org/10.1364/OL.31.001914} {\bibfield
  {journal} {\bibinfo  {journal} {Opt. Lett.}\ }\textbf {\bibinfo {volume}
  {31}},\ \bibinfo {pages} {1914} (\bibinfo {year} {2006})}\BibitemShut
  {NoStop}%
\bibitem [{\citenamefont {Cousin}\ \emph {et~al.}(2012)\citenamefont {Cousin},
  \citenamefont {Bueno}, \citenamefont {Forget}, \citenamefont {Austin},\ and\
  \citenamefont {Biegert}}]{Cousin:12}%
  \BibitemOpen
  \bibfield  {author} {\bibinfo {author} {\bibfnamefont {S.~L.}\ \bibnamefont
  {Cousin}}, \bibinfo {author} {\bibfnamefont {J.~M.}\ \bibnamefont {Bueno}},
  \bibinfo {author} {\bibfnamefont {N.}~\bibnamefont {Forget}}, \bibinfo
  {author} {\bibfnamefont {D.~R.}\ \bibnamefont {Austin}},\ and\ \bibinfo
  {author} {\bibfnamefont {J.}~\bibnamefont {Biegert}},\ }\bibfield  {title}
  {\bibinfo {title} {Three-dimensional spatiotemporal pulse characterization
  with an acousto-optic pulse shaper and a hartmann shack wavefront sensor},\
  }\href {https://doi.org/10.1364/OL.37.003291} {\bibfield  {journal} {\bibinfo
   {journal} {Opt. Lett.}\ }\textbf {\bibinfo {volume} {37}},\ \bibinfo {pages}
  {3291} (\bibinfo {year} {2012})}\BibitemShut {NoStop}%
\bibitem [{\citenamefont {Gabolde}\ and\ \citenamefont
  {Trebino}(2006)}]{Gabolde:06}%
  \BibitemOpen
  \bibfield  {author} {\bibinfo {author} {\bibfnamefont {P.}~\bibnamefont
  {Gabolde}}\ and\ \bibinfo {author} {\bibfnamefont {R.}~\bibnamefont
  {Trebino}},\ }\bibfield  {title} {\bibinfo {title} {Single-shot measurement
  of the full spatio-temporal field of ultrashort pulses with multi-spectral
  digital holography},\ }\href {https://doi.org/10.1364/OE.14.011460}
  {\bibfield  {journal} {\bibinfo  {journal} {Opt. Express}\ }\textbf {\bibinfo
  {volume} {14}},\ \bibinfo {pages} {11460} (\bibinfo {year}
  {2006})}\BibitemShut {NoStop}%
\bibitem [{\citenamefont {Miranda}\ \emph {et~al.}(2014)\citenamefont
  {Miranda}, \citenamefont {Kotur}, \citenamefont {Rudawski}, \citenamefont
  {Guo}, \citenamefont {Harth}, \citenamefont {L'Huillier},\ and\ \citenamefont
  {Arnold}}]{Miranda:14}%
  \BibitemOpen
  \bibfield  {author} {\bibinfo {author} {\bibfnamefont {M.}~\bibnamefont
  {Miranda}}, \bibinfo {author} {\bibfnamefont {M.}~\bibnamefont {Kotur}},
  \bibinfo {author} {\bibfnamefont {P.}~\bibnamefont {Rudawski}}, \bibinfo
  {author} {\bibfnamefont {C.}~\bibnamefont {Guo}}, \bibinfo {author}
  {\bibfnamefont {A.}~\bibnamefont {Harth}}, \bibinfo {author} {\bibfnamefont
  {A.}~\bibnamefont {L'Huillier}},\ and\ \bibinfo {author} {\bibfnamefont
  {C.~L.}\ \bibnamefont {Arnold}},\ }\bibfield  {title} {\bibinfo {title}
  {Spatiotemporal characterization of ultrashort laser pulses using spatially
  resolved fourier transform spectrometry},\ }\href
  {https://doi.org/10.1364/OL.39.005142} {\bibfield  {journal} {\bibinfo
  {journal} {Opt. Lett.}\ }\textbf {\bibinfo {volume} {39}},\ \bibinfo {pages}
  {5142} (\bibinfo {year} {2014})}\BibitemShut {NoStop}%
\bibitem [{\citenamefont {Pariente}\ \emph {et~al.}(2016)\citenamefont
  {Pariente}, \citenamefont {Gallet}, \citenamefont {Borot}, \citenamefont
  {Gobert},\ and\ \citenamefont {Qu{\'e}r{\'e}}}]{pariente2016space}%
  \BibitemOpen
  \bibfield  {author} {\bibinfo {author} {\bibfnamefont {G.}~\bibnamefont
  {Pariente}}, \bibinfo {author} {\bibfnamefont {V.}~\bibnamefont {Gallet}},
  \bibinfo {author} {\bibfnamefont {A.}~\bibnamefont {Borot}}, \bibinfo
  {author} {\bibfnamefont {O.}~\bibnamefont {Gobert}},\ and\ \bibinfo {author}
  {\bibfnamefont {F.}~\bibnamefont {Qu{\'e}r{\'e}}},\ }\bibfield  {title}
  {\bibinfo {title} {Space-time characterization of ultra-intense femtosecond
  laser beams},\ }\href@noop {} {\bibfield  {journal} {\bibinfo  {journal}
  {Nature Photonics}\ }\textbf {\bibinfo {volume} {10}},\ \bibinfo {pages}
  {547} (\bibinfo {year} {2016})}\BibitemShut {NoStop}%
\bibitem [{\citenamefont {Moh}\ \emph {et~al.}(2006)\citenamefont {Moh},
  \citenamefont {Yuan}, \citenamefont {Bu}, \citenamefont {Low},\ and\
  \citenamefont {Burge}}]{moh2006direct}%
  \BibitemOpen
  \bibfield  {author} {\bibinfo {author} {\bibfnamefont {K.}~\bibnamefont
  {Moh}}, \bibinfo {author} {\bibfnamefont {X.-C.}\ \bibnamefont {Yuan}},
  \bibinfo {author} {\bibfnamefont {J.}~\bibnamefont {Bu}}, \bibinfo {author}
  {\bibfnamefont {D.}~\bibnamefont {Low}},\ and\ \bibinfo {author}
  {\bibfnamefont {R.}~\bibnamefont {Burge}},\ }\bibfield  {title} {\bibinfo
  {title} {Direct noninterference cylindrical vector beam generation applied in
  the femtosecond regime},\ }\href@noop {} {\bibfield  {journal} {\bibinfo
  {journal} {Applied physics letters}\ }\textbf {\bibinfo {volume} {89}},\
  \bibinfo {pages} {251114} (\bibinfo {year} {2006})}\BibitemShut {NoStop}%
\bibitem [{\citenamefont {Kong}\ \emph {et~al.}(2019)\citenamefont {Kong},
  \citenamefont {Larocque}, \citenamefont {Karimi}, \citenamefont {Corkum},\
  and\ \citenamefont {Zhang}}]{Kong:19}%
  \BibitemOpen
  \bibfield  {author} {\bibinfo {author} {\bibfnamefont {F.}~\bibnamefont
  {Kong}}, \bibinfo {author} {\bibfnamefont {H.}~\bibnamefont {Larocque}},
  \bibinfo {author} {\bibfnamefont {E.}~\bibnamefont {Karimi}}, \bibinfo
  {author} {\bibfnamefont {P.~B.}\ \bibnamefont {Corkum}},\ and\ \bibinfo
  {author} {\bibfnamefont {C.}~\bibnamefont {Zhang}},\ }\bibfield  {title}
  {\bibinfo {title} {Generating few-cycle radially polarized pulses},\ }\href
  {https://doi.org/10.1364/OPTICA.6.000160} {\bibfield  {journal} {\bibinfo
  {journal} {Optica}\ }\textbf {\bibinfo {volume} {6}},\ \bibinfo {pages} {160}
  (\bibinfo {year} {2019})}\BibitemShut {NoStop}%
\bibitem [{\citenamefont {Carbajo}\ \emph {et~al.}(2014)\citenamefont
  {Carbajo}, \citenamefont {Granados}, \citenamefont {Schimpf}, \citenamefont
  {Sell}, \citenamefont {Hong}, \citenamefont {Moses},\ and\ \citenamefont
  {K\"{a}rtner}}]{Carbajo:14}%
  \BibitemOpen
  \bibfield  {author} {\bibinfo {author} {\bibfnamefont {S.}~\bibnamefont
  {Carbajo}}, \bibinfo {author} {\bibfnamefont {E.}~\bibnamefont {Granados}},
  \bibinfo {author} {\bibfnamefont {D.}~\bibnamefont {Schimpf}}, \bibinfo
  {author} {\bibfnamefont {A.}~\bibnamefont {Sell}}, \bibinfo {author}
  {\bibfnamefont {K.-H.}\ \bibnamefont {Hong}}, \bibinfo {author}
  {\bibfnamefont {J.}~\bibnamefont {Moses}},\ and\ \bibinfo {author}
  {\bibfnamefont {F.~X.}\ \bibnamefont {K\"{a}rtner}},\ }\bibfield  {title}
  {\bibinfo {title} {Efficient generation of ultra-intense few-cycle radially
  polarized laser pulses},\ }\href {https://doi.org/10.1364/OL.39.002487}
  {\bibfield  {journal} {\bibinfo  {journal} {Opt. Lett.}\ }\textbf {\bibinfo
  {volume} {39}},\ \bibinfo {pages} {2487} (\bibinfo {year}
  {2014})}\BibitemShut {NoStop}%
\bibitem [{\citenamefont {Papasimakis}\ \emph {et~al.}(2018)\citenamefont
  {Papasimakis}, \citenamefont {Raybould}, \citenamefont {Fedotov},
  \citenamefont {Tsai}, \citenamefont {Youngs},\ and\ \citenamefont
  {Zheludev}}]{PhysRevB.97.201409}%
  \BibitemOpen
  \bibfield  {author} {\bibinfo {author} {\bibfnamefont {N.}~\bibnamefont
  {Papasimakis}}, \bibinfo {author} {\bibfnamefont {T.}~\bibnamefont
  {Raybould}}, \bibinfo {author} {\bibfnamefont {V.~A.}\ \bibnamefont
  {Fedotov}}, \bibinfo {author} {\bibfnamefont {D.~P.}\ \bibnamefont {Tsai}},
  \bibinfo {author} {\bibfnamefont {I.}~\bibnamefont {Youngs}},\ and\ \bibinfo
  {author} {\bibfnamefont {N.~I.}\ \bibnamefont {Zheludev}},\ }\bibfield
  {title} {\bibinfo {title} {Pulse generation scheme for flying electromagnetic
  doughnuts},\ }\href {https://doi.org/10.1103/PhysRevB.97.201409} {\bibfield
  {journal} {\bibinfo  {journal} {Phys. Rev. B}\ }\textbf {\bibinfo {volume}
  {97}},\ \bibinfo {pages} {201409} (\bibinfo {year} {2018})}\BibitemShut
  {NoStop}%
\bibitem [{\citenamefont {Trebino}\ and\ \citenamefont
  {Kane}(1993)}]{Trebino:93}%
  \BibitemOpen
  \bibfield  {author} {\bibinfo {author} {\bibfnamefont {R.}~\bibnamefont
  {Trebino}}\ and\ \bibinfo {author} {\bibfnamefont {D.~J.}\ \bibnamefont
  {Kane}},\ }\bibfield  {title} {\bibinfo {title} {Using phase retrieval to
  measure the intensity and phase of ultrashort pulses: frequency-resolved
  optical gating},\ }\href {https://doi.org/10.1364/JOSAA.10.001101} {\bibfield
   {journal} {\bibinfo  {journal} {J. Opt. Soc. Am. A}\ }\textbf {\bibinfo
  {volume} {10}},\ \bibinfo {pages} {1101} (\bibinfo {year}
  {1993})}\BibitemShut {NoStop}%
\bibitem [{\citenamefont {Iaconis}\ and\ \citenamefont
  {Walmsley}(1998)}]{Iaconis:98}%
  \BibitemOpen
  \bibfield  {author} {\bibinfo {author} {\bibfnamefont {C.}~\bibnamefont
  {Iaconis}}\ and\ \bibinfo {author} {\bibfnamefont {I.~A.}\ \bibnamefont
  {Walmsley}},\ }\bibfield  {title} {\bibinfo {title} {Spectral phase
  interferometry for direct electric-field reconstruction of ultrashort optical
  pulses},\ }\href {https://doi.org/10.1364/OL.23.000792} {\bibfield  {journal}
  {\bibinfo  {journal} {Opt. Lett.}\ }\textbf {\bibinfo {volume} {23}},\
  \bibinfo {pages} {792} (\bibinfo {year} {1998})}\BibitemShut {NoStop}%
\bibitem [{\citenamefont {Zdagkas}\ \emph {et~al.}(2020)\citenamefont
  {Zdagkas}, \citenamefont {Papasimakis}, \citenamefont {Savinov},\ and\
  \citenamefont {Zheludev}}]{zdagkas2020space}%
  \BibitemOpen
  \bibfield  {author} {\bibinfo {author} {\bibfnamefont {A.}~\bibnamefont
  {Zdagkas}}, \bibinfo {author} {\bibfnamefont {N.}~\bibnamefont
  {Papasimakis}}, \bibinfo {author} {\bibfnamefont {V.}~\bibnamefont
  {Savinov}},\ and\ \bibinfo {author} {\bibfnamefont {N.~I.}\ \bibnamefont
  {Zheludev}},\ }\bibfield  {title} {\bibinfo {title} {Space-time nonseparable
  pulses: Constructing isodiffracting donut pulses from plane waves and
  single-cycle pulses},\ }\href@noop {} {\bibfield  {journal} {\bibinfo
  {journal} {Physical Review A}\ }\textbf {\bibinfo {volume} {102}},\ \bibinfo
  {pages} {063512} (\bibinfo {year} {2020})}\BibitemShut {NoStop}%
\bibitem [{\citenamefont {Strickland}\ and\ \citenamefont
  {Mourou}(1985)}]{STRICKLAND1985447}%
  \BibitemOpen
  \bibfield  {author} {\bibinfo {author} {\bibfnamefont {D.}~\bibnamefont
  {Strickland}}\ and\ \bibinfo {author} {\bibfnamefont {G.}~\bibnamefont
  {Mourou}},\ }\bibfield  {title} {\bibinfo {title} {Compression of amplified
  chirped optical pulses},\ }\href
  {https://doi.org/https://doi.org/10.1016/0030-4018(85)90151-8} {\bibfield
  {journal} {\bibinfo  {journal} {Optics Communications}\ }\textbf {\bibinfo
  {volume} {55}},\ \bibinfo {pages} {447 } (\bibinfo {year}
  {1985})}\BibitemShut {NoStop}%
\bibitem [{\citenamefont {Pretzler}\ \emph
  {et~al.}(2000{\natexlab{b}})\citenamefont {Pretzler}, \citenamefont
  {Kasper},\ and\ \citenamefont {Witte}}]{pft_CPA_lasers}%
  \BibitemOpen
  \bibfield  {author} {\bibinfo {author} {\bibfnamefont {G.}~\bibnamefont
  {Pretzler}}, \bibinfo {author} {\bibfnamefont {A.}~\bibnamefont {Kasper}},\
  and\ \bibinfo {author} {\bibfnamefont {K.}~\bibnamefont {Witte}},\ }\bibfield
   {title} {\bibinfo {title} {Angular chirp and tilted light pulses in cpa
  lasers},\ }\href {https://doi.org/10.1007/s003400050001} {\bibfield
  {journal} {\bibinfo  {journal} {Applied Physics B: Lasers and Optics}\
  }\textbf {\bibinfo {volume} {70}},\ \bibinfo {pages} {1} (\bibinfo {year}
  {2000}{\natexlab{b}})}\BibitemShut {NoStop}%
\bibitem [{\citenamefont {Alonso}\ \emph {et~al.}(2020)\citenamefont {Alonso},
  \citenamefont {Lopez-Quintas}, \citenamefont {Holgado}, \citenamefont
  {Drevinskas}, \citenamefont {Kazansky}, \citenamefont
  {Hern{\'a}ndez-Garc{\'\i}a},\ and\ \citenamefont
  {Sola}}]{alonso2020complete}%
  \BibitemOpen
  \bibfield  {author} {\bibinfo {author} {\bibfnamefont {B.}~\bibnamefont
  {Alonso}}, \bibinfo {author} {\bibfnamefont {I.}~\bibnamefont
  {Lopez-Quintas}}, \bibinfo {author} {\bibfnamefont {W.}~\bibnamefont
  {Holgado}}, \bibinfo {author} {\bibfnamefont {R.}~\bibnamefont {Drevinskas}},
  \bibinfo {author} {\bibfnamefont {P.~G.}\ \bibnamefont {Kazansky}}, \bibinfo
  {author} {\bibfnamefont {C.}~\bibnamefont {Hern{\'a}ndez-Garc{\'\i}a}},\ and\
  \bibinfo {author} {\bibfnamefont {{\'I}.~J.}\ \bibnamefont {Sola}},\
  }\bibfield  {title} {\bibinfo {title} {Complete spatiotemporal and
  polarization characterization of ultrafast vector beams},\ }\href@noop {}
  {\bibfield  {journal} {\bibinfo  {journal} {Communications Physics}\ }\textbf
  {\bibinfo {volume} {3}},\ \bibinfo {pages} {1} (\bibinfo {year}
  {2020})}\BibitemShut {NoStop}%
\bibitem [{\citenamefont {Zdagkas}(2021)}]{azdagkas/TERMITES_MAZE_python_pure}%
  \BibitemOpen
  \bibfield  {author} {\bibinfo {author} {\bibfnamefont {A.}~\bibnamefont
  {Zdagkas}},\ }\href {https://doi.org/10.5258/SOTON/D1678} {\bibinfo {title}
  {Termites-maze python module}} (\bibinfo {year} {2021})\BibitemShut {NoStop}%
\bibitem [{\citenamefont {Allen}\ \emph {et~al.}(1992)\citenamefont {Allen},
  \citenamefont {Beijersbergen}, \citenamefont {Spreeuw},\ and\ \citenamefont
  {Woerdman}}]{PhysRevA.45.8185}%
  \BibitemOpen
  \bibfield  {author} {\bibinfo {author} {\bibfnamefont {L.}~\bibnamefont
  {Allen}}, \bibinfo {author} {\bibfnamefont {M.~W.}\ \bibnamefont
  {Beijersbergen}}, \bibinfo {author} {\bibfnamefont {R.~J.~C.}\ \bibnamefont
  {Spreeuw}},\ and\ \bibinfo {author} {\bibfnamefont {J.~P.}\ \bibnamefont
  {Woerdman}},\ }\bibfield  {title} {\bibinfo {title} {Orbital angular momentum
  of light and the transformation of laguerre-gaussian laser modes},\ }\href
  {https://doi.org/10.1103/PhysRevA.45.8185} {\bibfield  {journal} {\bibinfo
  {journal} {Phys. Rev. A}\ }\textbf {\bibinfo {volume} {45}},\ \bibinfo
  {pages} {8185} (\bibinfo {year} {1992})}\BibitemShut {NoStop}%
\bibitem [{\citenamefont {Rubinsztein-Dunlop}\ \emph
  {et~al.}(2016)\citenamefont {Rubinsztein-Dunlop}, \citenamefont {Forbes},
  \citenamefont {Berry}, \citenamefont {Dennis}, \citenamefont {Andrews},
  \citenamefont {Mansuripur}, \citenamefont {Denz}, \citenamefont {Alpmann},
  \citenamefont {Banzer}, \citenamefont {Bauer}, \citenamefont {Karimi},
  \citenamefont {Marrucci}, \citenamefont {Padgett}, \citenamefont
  {Ritsch-Marte}, \citenamefont {Litchinitser}, \citenamefont {Bigelow},
  \citenamefont {Rosales-Guzm{\'{a}}n}, \citenamefont {Belmonte}, \citenamefont
  {Torres}, \citenamefont {Neely}, \citenamefont {Baker}, \citenamefont
  {Gordon}, \citenamefont {Stilgoe}, \citenamefont {Romero}, \citenamefont
  {White}, \citenamefont {Fickler}, \citenamefont {Willner}, \citenamefont
  {Xie}, \citenamefont {McMorran},\ and\ \citenamefont
  {Weiner}}]{Rubinsztein_Dunlop_2016}%
  \BibitemOpen
  \bibfield  {author} {\bibinfo {author} {\bibfnamefont {H.}~\bibnamefont
  {Rubinsztein-Dunlop}}, \bibinfo {author} {\bibfnamefont {A.}~\bibnamefont
  {Forbes}}, \bibinfo {author} {\bibfnamefont {M.~V.}\ \bibnamefont {Berry}},
  \bibinfo {author} {\bibfnamefont {M.~R.}\ \bibnamefont {Dennis}}, \bibinfo
  {author} {\bibfnamefont {D.~L.}\ \bibnamefont {Andrews}}, \bibinfo {author}
  {\bibfnamefont {M.}~\bibnamefont {Mansuripur}}, \bibinfo {author}
  {\bibfnamefont {C.}~\bibnamefont {Denz}}, \bibinfo {author} {\bibfnamefont
  {C.}~\bibnamefont {Alpmann}}, \bibinfo {author} {\bibfnamefont
  {P.}~\bibnamefont {Banzer}}, \bibinfo {author} {\bibfnamefont
  {T.}~\bibnamefont {Bauer}}, \bibinfo {author} {\bibfnamefont
  {E.}~\bibnamefont {Karimi}}, \bibinfo {author} {\bibfnamefont
  {L.}~\bibnamefont {Marrucci}}, \bibinfo {author} {\bibfnamefont
  {M.}~\bibnamefont {Padgett}}, \bibinfo {author} {\bibfnamefont
  {M.}~\bibnamefont {Ritsch-Marte}}, \bibinfo {author} {\bibfnamefont {N.~M.}\
  \bibnamefont {Litchinitser}}, \bibinfo {author} {\bibfnamefont {N.~P.}\
  \bibnamefont {Bigelow}}, \bibinfo {author} {\bibfnamefont {C.}~\bibnamefont
  {Rosales-Guzm{\'{a}}n}}, \bibinfo {author} {\bibfnamefont {A.}~\bibnamefont
  {Belmonte}}, \bibinfo {author} {\bibfnamefont {J.~P.}\ \bibnamefont
  {Torres}}, \bibinfo {author} {\bibfnamefont {T.~W.}\ \bibnamefont {Neely}},
  \bibinfo {author} {\bibfnamefont {M.}~\bibnamefont {Baker}}, \bibinfo
  {author} {\bibfnamefont {R.}~\bibnamefont {Gordon}}, \bibinfo {author}
  {\bibfnamefont {A.~B.}\ \bibnamefont {Stilgoe}}, \bibinfo {author}
  {\bibfnamefont {J.}~\bibnamefont {Romero}}, \bibinfo {author} {\bibfnamefont
  {A.~G.}\ \bibnamefont {White}}, \bibinfo {author} {\bibfnamefont
  {R.}~\bibnamefont {Fickler}}, \bibinfo {author} {\bibfnamefont {A.~E.}\
  \bibnamefont {Willner}}, \bibinfo {author} {\bibfnamefont {G.}~\bibnamefont
  {Xie}}, \bibinfo {author} {\bibfnamefont {B.}~\bibnamefont {McMorran}},\ and\
  \bibinfo {author} {\bibfnamefont {A.~M.}\ \bibnamefont {Weiner}},\ }\bibfield
   {title} {\bibinfo {title} {Roadmap on structured light},\ }\href
  {https://doi.org/10.1088/2040-8978/19/1/013001} {\bibfield  {journal}
  {\bibinfo  {journal} {Journal of Optics}\ }\textbf {\bibinfo {volume} {19}},\
  \bibinfo {pages} {013001} (\bibinfo {year} {2016})}\BibitemShut {NoStop}%
\bibitem [{\citenamefont {Porras}(2002)}]{PhysRevE.65.026606}%
  \BibitemOpen
  \bibfield  {author} {\bibinfo {author} {\bibfnamefont {M.~A.}\ \bibnamefont
  {Porras}},\ }\bibfield  {title} {\bibinfo {title} {Diffraction effects in
  few-cycle optical pulses},\ }\href
  {https://doi.org/10.1103/PhysRevE.65.026606} {\bibfield  {journal} {\bibinfo
  {journal} {Phys. Rev. E}\ }\textbf {\bibinfo {volume} {65}},\ \bibinfo
  {pages} {026606} (\bibinfo {year} {2002})}\BibitemShut {NoStop}%
\bibitem [{\citenamefont {Hoff}\ \emph {et~al.}(2017)\citenamefont {Hoff},
  \citenamefont {Krüger}, \citenamefont {Maisenbacher}, \citenamefont
  {Sayler}, \citenamefont {Paulus},\ and\ \citenamefont
  {Hommelhoff}}]{Hoff_nature}%
  \BibitemOpen
  \bibfield  {author} {\bibinfo {author} {\bibfnamefont {D.}~\bibnamefont
  {Hoff}}, \bibinfo {author} {\bibfnamefont {M.}~\bibnamefont {Krüger}},
  \bibinfo {author} {\bibfnamefont {L.}~\bibnamefont {Maisenbacher}}, \bibinfo
  {author} {\bibfnamefont {A.}~\bibnamefont {Sayler}}, \bibinfo {author}
  {\bibfnamefont {G.}~\bibnamefont {Paulus}},\ and\ \bibinfo {author}
  {\bibfnamefont {P.}~\bibnamefont {Hommelhoff}},\ }\bibfield  {title}
  {\bibinfo {title} {Tracing the phase of focused broadband laser pulses},\
  }\href {https://doi.org/10.1038/nphys4185} {\bibfield  {journal} {\bibinfo
  {journal} {Nature Physics}\ }\textbf {\bibinfo {volume} {13}} (\bibinfo
  {year} {2017})}\BibitemShut {NoStop}%
\bibitem [{\citenamefont {Zhang}\ \emph {et~al.}(2020)\citenamefont {Zhang},
  \citenamefont {Zille}, \citenamefont {Hoff}, \citenamefont {Wustelt},
  \citenamefont {W\"urzler}, \citenamefont {M\"oller}, \citenamefont {Sayler},\
  and\ \citenamefont {Paulus}}]{PhysRevLett.124.133202}%
  \BibitemOpen
  \bibfield  {author} {\bibinfo {author} {\bibfnamefont {Y.}~\bibnamefont
  {Zhang}}, \bibinfo {author} {\bibfnamefont {D.}~\bibnamefont {Zille}},
  \bibinfo {author} {\bibfnamefont {D.}~\bibnamefont {Hoff}}, \bibinfo {author}
  {\bibfnamefont {P.}~\bibnamefont {Wustelt}}, \bibinfo {author} {\bibfnamefont
  {D.}~\bibnamefont {W\"urzler}}, \bibinfo {author} {\bibfnamefont
  {M.}~\bibnamefont {M\"oller}}, \bibinfo {author} {\bibfnamefont {A.~M.}\
  \bibnamefont {Sayler}},\ and\ \bibinfo {author} {\bibfnamefont {G.~G.}\
  \bibnamefont {Paulus}},\ }\bibfield  {title} {\bibinfo {title} {Observing the
  importance of the phase-volume effect for few-cycle light-matter
  interactions},\ }\href {https://doi.org/10.1103/PhysRevLett.124.133202}
  {\bibfield  {journal} {\bibinfo  {journal} {Phys. Rev. Lett.}\ }\textbf
  {\bibinfo {volume} {124}},\ \bibinfo {pages} {133202} (\bibinfo {year}
  {2020})}\BibitemShut {NoStop}%
\bibitem [{\citenamefont {Forbes}\ \emph {et~al.}(2019)\citenamefont {Forbes},
  \citenamefont {Aiello},\ and\ \citenamefont
  {Ndagano}}]{forbes2019classically}%
  \BibitemOpen
  \bibfield  {author} {\bibinfo {author} {\bibfnamefont {A.}~\bibnamefont
  {Forbes}}, \bibinfo {author} {\bibfnamefont {A.}~\bibnamefont {Aiello}},\
  and\ \bibinfo {author} {\bibfnamefont {B.}~\bibnamefont {Ndagano}},\
  }\bibfield  {title} {\bibinfo {title} {Classically entangled light},\ }in\
  \href@noop {} {\emph {\bibinfo {booktitle} {Progress in Optics}}},\
  Vol.~\bibinfo {volume} {64}\ (\bibinfo  {publisher} {Elsevier},\ \bibinfo
  {year} {2019})\ pp.\ \bibinfo {pages} {99--153}\BibitemShut {NoStop}%
\bibitem [{\citenamefont {Zhan}(2009)}]{Zhan:09}%
  \BibitemOpen
  \bibfield  {author} {\bibinfo {author} {\bibfnamefont {Q.}~\bibnamefont
  {Zhan}},\ }\bibfield  {title} {\bibinfo {title} {Cylindrical vector beams:
  from mathematical concepts to applications},\ }\href
  {https://doi.org/10.1364/AOP.1.000001} {\bibfield  {journal} {\bibinfo
  {journal} {Adv. Opt. Photon.}\ }\textbf {\bibinfo {volume} {1}},\ \bibinfo
  {pages} {1} (\bibinfo {year} {2009})}\BibitemShut {NoStop}%
\bibitem [{\citenamefont {Dorn}\ \emph {et~al.}(2003)\citenamefont {Dorn},
  \citenamefont {Quabis},\ and\ \citenamefont
  {Leuchs}}]{PhysRevLett.91.233901}%
  \BibitemOpen
  \bibfield  {author} {\bibinfo {author} {\bibfnamefont {R.}~\bibnamefont
  {Dorn}}, \bibinfo {author} {\bibfnamefont {S.}~\bibnamefont {Quabis}},\ and\
  \bibinfo {author} {\bibfnamefont {G.}~\bibnamefont {Leuchs}},\ }\bibfield
  {title} {\bibinfo {title} {Sharper focus for a radially polarized light
  beam},\ }\href {https://doi.org/10.1103/PhysRevLett.91.233901} {\bibfield
  {journal} {\bibinfo  {journal} {Phys. Rev. Lett.}\ }\textbf {\bibinfo
  {volume} {91}},\ \bibinfo {pages} {233901} (\bibinfo {year}
  {2003})}\BibitemShut {NoStop}%
\bibitem [{\citenamefont {Michihata}\ \emph {et~al.}(2009)\citenamefont
  {Michihata}, \citenamefont {Hayashi},\ and\ \citenamefont
  {Takaya}}]{Michihata:09}%
  \BibitemOpen
  \bibfield  {author} {\bibinfo {author} {\bibfnamefont {M.}~\bibnamefont
  {Michihata}}, \bibinfo {author} {\bibfnamefont {T.}~\bibnamefont {Hayashi}},\
  and\ \bibinfo {author} {\bibfnamefont {Y.}~\bibnamefont {Takaya}},\
  }\bibfield  {title} {\bibinfo {title} {Measurement of axial and transverse
  trapping stiffness of optical tweezers in air using a radially polarized
  beam},\ }\href {https://doi.org/10.1364/AO.48.006143} {\bibfield  {journal}
  {\bibinfo  {journal} {Appl. Opt.}\ }\textbf {\bibinfo {volume} {48}},\
  \bibinfo {pages} {6143} (\bibinfo {year} {2009})}\BibitemShut {NoStop}%
\bibitem [{\citenamefont {Moradi}\ \emph {et~al.}(2019)\citenamefont {Moradi},
  \citenamefont {Shahabadi}, \citenamefont {Madadi}, \citenamefont {Karimi},\
  and\ \citenamefont {Hajizadeh}}]{Moradi:19}%
  \BibitemOpen
  \bibfield  {author} {\bibinfo {author} {\bibfnamefont {H.}~\bibnamefont
  {Moradi}}, \bibinfo {author} {\bibfnamefont {V.}~\bibnamefont {Shahabadi}},
  \bibinfo {author} {\bibfnamefont {E.}~\bibnamefont {Madadi}}, \bibinfo
  {author} {\bibfnamefont {E.}~\bibnamefont {Karimi}},\ and\ \bibinfo {author}
  {\bibfnamefont {F.}~\bibnamefont {Hajizadeh}},\ }\bibfield  {title} {\bibinfo
  {title} {Efficient optical trapping with cylindrical vector beams},\ }\href
  {https://doi.org/10.1364/OE.27.007266} {\bibfield  {journal} {\bibinfo
  {journal} {Opt. Express}\ }\textbf {\bibinfo {volume} {27}},\ \bibinfo
  {pages} {7266} (\bibinfo {year} {2019})}\BibitemShut {NoStop}%
\bibitem [{\citenamefont {Chen}\ \emph {et~al.}(2013)\citenamefont {Chen},
  \citenamefont {Agarwal}, \citenamefont {Sheppard},\ and\ \citenamefont
  {Chen}}]{Chen:13}%
  \BibitemOpen
  \bibfield  {author} {\bibinfo {author} {\bibfnamefont {R.}~\bibnamefont
  {Chen}}, \bibinfo {author} {\bibfnamefont {K.}~\bibnamefont {Agarwal}},
  \bibinfo {author} {\bibfnamefont {C.~J.~R.}\ \bibnamefont {Sheppard}},\ and\
  \bibinfo {author} {\bibfnamefont {X.}~\bibnamefont {Chen}},\ }\bibfield
  {title} {\bibinfo {title} {Imaging using cylindrical vector beams in a
  high-numerical-aperture microscopy system},\ }\href
  {https://doi.org/10.1364/OL.38.003111} {\bibfield  {journal} {\bibinfo
  {journal} {Opt. Lett.}\ }\textbf {\bibinfo {volume} {38}},\ \bibinfo {pages}
  {3111} (\bibinfo {year} {2013})}\BibitemShut {NoStop}%
\bibitem [{\citenamefont {Li}\ \emph {et~al.}(2011)\citenamefont {Li},
  \citenamefont {Cao},\ and\ \citenamefont {Gu}}]{Li:11}%
  \BibitemOpen
  \bibfield  {author} {\bibinfo {author} {\bibfnamefont {X.}~\bibnamefont
  {Li}}, \bibinfo {author} {\bibfnamefont {Y.}~\bibnamefont {Cao}},\ and\
  \bibinfo {author} {\bibfnamefont {M.}~\bibnamefont {Gu}},\ }\bibfield
  {title} {\bibinfo {title} {Superresolution-focal-volume induced 3.0
  tbytes/disk capacity by focusing a radially polarized beam},\ }\href
  {https://doi.org/10.1364/OL.36.002510} {\bibfield  {journal} {\bibinfo
  {journal} {Opt. Lett.}\ }\textbf {\bibinfo {volume} {36}},\ \bibinfo {pages}
  {2510} (\bibinfo {year} {2011})}\BibitemShut {NoStop}%
\bibitem [{\citenamefont {Milione}\ \emph {et~al.}(2015)\citenamefont
  {Milione}, \citenamefont {Nguyen}, \citenamefont {Leach}, \citenamefont
  {Nolan},\ and\ \citenamefont {Alfano}}]{Milione:15}%
  \BibitemOpen
  \bibfield  {author} {\bibinfo {author} {\bibfnamefont {G.}~\bibnamefont
  {Milione}}, \bibinfo {author} {\bibfnamefont {T.~A.}\ \bibnamefont {Nguyen}},
  \bibinfo {author} {\bibfnamefont {J.}~\bibnamefont {Leach}}, \bibinfo
  {author} {\bibfnamefont {D.~A.}\ \bibnamefont {Nolan}},\ and\ \bibinfo
  {author} {\bibfnamefont {R.~R.}\ \bibnamefont {Alfano}},\ }\bibfield  {title}
  {\bibinfo {title} {Using the nonseparability of vector beams to encode
  information for optical communication},\ }\href
  {https://doi.org/10.1364/OL.40.004887} {\bibfield  {journal} {\bibinfo
  {journal} {Opt. Lett.}\ }\textbf {\bibinfo {volume} {40}},\ \bibinfo {pages}
  {4887} (\bibinfo {year} {2015})}\BibitemShut {NoStop}%
\bibitem [{\citenamefont {Konrad}\ and\ \citenamefont
  {Forbes}(2019)}]{doi:10.1080/00107514.2019.1580433}%
  \BibitemOpen
  \bibfield  {author} {\bibinfo {author} {\bibfnamefont {T.}~\bibnamefont
  {Konrad}}\ and\ \bibinfo {author} {\bibfnamefont {A.}~\bibnamefont
  {Forbes}},\ }\bibfield  {title} {\bibinfo {title} {Quantum mechanics and
  classical light},\ }\href {https://doi.org/10.1080/00107514.2019.1580433}
  {\bibfield  {journal} {\bibinfo  {journal} {Contemporary Physics}\ }\textbf
  {\bibinfo {volume} {60}},\ \bibinfo {pages} {1} (\bibinfo {year} {2019})},\
  \Eprint {https://arxiv.org/abs/https://doi.org/10.1080/00107514.2019.1580433}
  {https://doi.org/10.1080/00107514.2019.1580433} \BibitemShut {NoStop}%
\bibitem [{\citenamefont {McLaren}\ \emph {et~al.}(2015)\citenamefont
  {McLaren}, \citenamefont {Konrad},\ and\ \citenamefont
  {Forbes}}]{mclaren2015measuring}%
  \BibitemOpen
  \bibfield  {author} {\bibinfo {author} {\bibfnamefont {M.}~\bibnamefont
  {McLaren}}, \bibinfo {author} {\bibfnamefont {T.}~\bibnamefont {Konrad}},\
  and\ \bibinfo {author} {\bibfnamefont {A.}~\bibnamefont {Forbes}},\
  }\bibfield  {title} {\bibinfo {title} {Measuring the nonseparability of
  vector vortex beams},\ }\href@noop {} {\bibfield  {journal} {\bibinfo
  {journal} {Physical Review A}\ }\textbf {\bibinfo {volume} {92}},\ \bibinfo
  {pages} {023833} (\bibinfo {year} {2015})}\BibitemShut {NoStop}%
\bibitem [{\citenamefont {Allegre}\ \emph {et~al.}(2012)\citenamefont
  {Allegre}, \citenamefont {Perrie}, \citenamefont {Edwardson}, \citenamefont
  {Dearden},\ and\ \citenamefont {Watkins}}]{Allegre_2012}%
  \BibitemOpen
  \bibfield  {author} {\bibinfo {author} {\bibfnamefont {O.~J.}\ \bibnamefont
  {Allegre}}, \bibinfo {author} {\bibfnamefont {W.}~\bibnamefont {Perrie}},
  \bibinfo {author} {\bibfnamefont {S.~P.}\ \bibnamefont {Edwardson}}, \bibinfo
  {author} {\bibfnamefont {G.}~\bibnamefont {Dearden}},\ and\ \bibinfo {author}
  {\bibfnamefont {K.~G.}\ \bibnamefont {Watkins}},\ }\bibfield  {title}
  {\bibinfo {title} {Laser microprocessing of steel with radially and
  azimuthally polarized femtosecond vortex pulses},\ }\href
  {https://doi.org/10.1088/2040-8978/14/8/085601} {\bibfield  {journal}
  {\bibinfo  {journal} {Journal of Optics}\ }\textbf {\bibinfo {volume} {14}},\
  \bibinfo {pages} {085601} (\bibinfo {year} {2012})}\BibitemShut {NoStop}%
\bibitem [{\citenamefont {Hellwarth}\ and\ \citenamefont
  {Nouchi}(1996)}]{PhysRevE.54.889}%
  \BibitemOpen
  \bibfield  {author} {\bibinfo {author} {\bibfnamefont {R.~W.}\ \bibnamefont
  {Hellwarth}}\ and\ \bibinfo {author} {\bibfnamefont {P.}~\bibnamefont
  {Nouchi}},\ }\bibfield  {title} {\bibinfo {title} {Focused one-cycle
  electromagnetic pulses},\ }\href {https://doi.org/10.1103/PhysRevE.54.889}
  {\bibfield  {journal} {\bibinfo  {journal} {Phys. Rev. E}\ }\textbf {\bibinfo
  {volume} {54}},\ \bibinfo {pages} {889} (\bibinfo {year} {1996})}\BibitemShut
  {NoStop}%
\bibitem [{\citenamefont {Zdagkas}\ \emph {et~al.}(2019)\citenamefont
  {Zdagkas}, \citenamefont {Papasimakis}, \citenamefont {Savinov},
  \citenamefont {Dennis},\ and\ \citenamefont
  {Zheludev}}]{zdagkas2019singularities}%
  \BibitemOpen
  \bibfield  {author} {\bibinfo {author} {\bibfnamefont {A.}~\bibnamefont
  {Zdagkas}}, \bibinfo {author} {\bibfnamefont {N.}~\bibnamefont
  {Papasimakis}}, \bibinfo {author} {\bibfnamefont {V.}~\bibnamefont
  {Savinov}}, \bibinfo {author} {\bibfnamefont {M.~R.}\ \bibnamefont
  {Dennis}},\ and\ \bibinfo {author} {\bibfnamefont {N.~I.}\ \bibnamefont
  {Zheludev}},\ }\bibfield  {title} {\bibinfo {title} {Singularities in the
  flying electromagnetic doughnuts},\ }\href@noop {} {\bibfield  {journal}
  {\bibinfo  {journal} {Nanophotonics}\ }\textbf {\bibinfo {volume} {8}},\
  \bibinfo {pages} {1379} (\bibinfo {year} {2019})}\BibitemShut {NoStop}%
\bibitem [{\citenamefont {Raybould}\ \emph {et~al.}(2016)\citenamefont
  {Raybould}, \citenamefont {Fedotov}, \citenamefont {Papasimakis},
  \citenamefont {Youngs},\ and\ \citenamefont
  {Zheludev}}]{Raybould_interaction}%
  \BibitemOpen
  \bibfield  {author} {\bibinfo {author} {\bibfnamefont {T.}~\bibnamefont
  {Raybould}}, \bibinfo {author} {\bibfnamefont {V.}~\bibnamefont {Fedotov}},
  \bibinfo {author} {\bibfnamefont {N.}~\bibnamefont {Papasimakis}}, \bibinfo
  {author} {\bibfnamefont {I.}~\bibnamefont {Youngs}},\ and\ \bibinfo {author}
  {\bibfnamefont {N.}~\bibnamefont {Zheludev}},\ }\bibfield  {title} {\bibinfo
  {title} {Focused electromagnetic doughnut pulses and their interaction with
  interfaces and nanostructures},\ }\href
  {https://doi.org/10.1364/OE.24.003150} {\bibfield  {journal} {\bibinfo
  {journal} {Opt. Express}\ }\textbf {\bibinfo {volume} {24}},\ \bibinfo
  {pages} {3150} (\bibinfo {year} {2016})}\BibitemShut {NoStop}%
\bibitem [{\citenamefont {Raybould}\ \emph {et~al.}(2017)\citenamefont
  {Raybould}, \citenamefont {Fedotov}, \citenamefont {Papasimakis},
  \citenamefont {Youngs},\ and\ \citenamefont
  {Zheludev}}]{Raybould_anapole_excitation}%
  \BibitemOpen
  \bibfield  {author} {\bibinfo {author} {\bibfnamefont {T.}~\bibnamefont
  {Raybould}}, \bibinfo {author} {\bibfnamefont {V.~A.}\ \bibnamefont
  {Fedotov}}, \bibinfo {author} {\bibfnamefont {N.}~\bibnamefont
  {Papasimakis}}, \bibinfo {author} {\bibfnamefont {I.}~\bibnamefont
  {Youngs}},\ and\ \bibinfo {author} {\bibfnamefont {N.~I.}\ \bibnamefont
  {Zheludev}},\ }\bibfield  {title} {\bibinfo {title} {Exciting dynamic
  anapoles with electromagnetic doughnut pulses},\ }\href
  {https://doi.org/10.1063/1.4999368} {\bibfield  {journal} {\bibinfo
  {journal} {Applied Physics Letters}\ }\textbf {\bibinfo {volume} {111}},\
  \bibinfo {pages} {081104} (\bibinfo {year} {2017})},\ \Eprint
  {https://arxiv.org/abs/http://dx.doi.org/10.1063/1.4999368}
  {http://dx.doi.org/10.1063/1.4999368} \BibitemShut {NoStop}%
\bibitem [{\citenamefont {Lozovoy}\ \emph {et~al.}(2004)\citenamefont
  {Lozovoy}, \citenamefont {Pastirk},\ and\ \citenamefont
  {Dantus}}]{Lozovoy:04}%
  \BibitemOpen
  \bibfield  {author} {\bibinfo {author} {\bibfnamefont {V.~V.}\ \bibnamefont
  {Lozovoy}}, \bibinfo {author} {\bibfnamefont {I.}~\bibnamefont {Pastirk}},\
  and\ \bibinfo {author} {\bibfnamefont {M.}~\bibnamefont {Dantus}},\
  }\bibfield  {title} {\bibinfo {title} {Multiphoton intrapulse interference.
  iv. ultrashort laser pulse spectral phase characterization and
  compensation},\ }\href {https://doi.org/10.1364/OL.29.000775} {\bibfield
  {journal} {\bibinfo  {journal} {Opt. Lett.}\ }\textbf {\bibinfo {volume}
  {29}},\ \bibinfo {pages} {775} (\bibinfo {year} {2004})}\BibitemShut
  {NoStop}%
\bibitem [{\citenamefont {Shannon}(1949)}]{shannon1949communication}%
  \BibitemOpen
  \bibfield  {author} {\bibinfo {author} {\bibfnamefont {C.~E.}\ \bibnamefont
  {Shannon}},\ }\bibfield  {title} {\bibinfo {title} {Communication in the
  presence of noise},\ }\href@noop {} {\bibfield  {journal} {\bibinfo
  {journal} {Proceedings of the IRE}\ }\textbf {\bibinfo {volume} {37}},\
  \bibinfo {pages} {10} (\bibinfo {year} {1949})}\BibitemShut {NoStop}%
\bibitem [{\citenamefont {Machavariani}\ \emph {et~al.}(2007)\citenamefont
  {Machavariani}, \citenamefont {Lumer}, \citenamefont {Moshe}, \citenamefont
  {Meir},\ and\ \citenamefont {Jackel}}]{Machavariani:07}%
  \BibitemOpen
  \bibfield  {author} {\bibinfo {author} {\bibfnamefont {G.}~\bibnamefont
  {Machavariani}}, \bibinfo {author} {\bibfnamefont {Y.}~\bibnamefont {Lumer}},
  \bibinfo {author} {\bibfnamefont {I.}~\bibnamefont {Moshe}}, \bibinfo
  {author} {\bibfnamefont {A.}~\bibnamefont {Meir}},\ and\ \bibinfo {author}
  {\bibfnamefont {S.}~\bibnamefont {Jackel}},\ }\bibfield  {title} {\bibinfo
  {title} {Efficient extracavity generation of radially and azimuthally
  polarized beams},\ }\href {https://doi.org/10.1364/OL.32.001468} {\bibfield
  {journal} {\bibinfo  {journal} {Opt. Lett.}\ }\textbf {\bibinfo {volume}
  {32}},\ \bibinfo {pages} {1468} (\bibinfo {year} {2007})}\BibitemShut
  {NoStop}%
\bibitem [{\citenamefont {Jolly}\ \emph {et~al.}(2020)\citenamefont {Jolly},
  \citenamefont {Gobert},\ and\ \citenamefont
  {Qu{\'e}r{\'e}}}]{jolly2020spatio}%
  \BibitemOpen
  \bibfield  {author} {\bibinfo {author} {\bibfnamefont {S.~W.}\ \bibnamefont
  {Jolly}}, \bibinfo {author} {\bibfnamefont {O.}~\bibnamefont {Gobert}},\ and\
  \bibinfo {author} {\bibfnamefont {F.}~\bibnamefont {Qu{\'e}r{\'e}}},\
  }\bibfield  {title} {\bibinfo {title} {Spatio-temporal characterization of
  ultrashort laser beams: a tutorial},\ }\href@noop {} {\bibfield  {journal}
  {\bibinfo  {journal} {Journal of Optics}\ }\textbf {\bibinfo {volume} {22}},\
  \bibinfo {pages} {103501} (\bibinfo {year} {2020})}\BibitemShut {NoStop}%
\bibitem [{\citenamefont {Sola}\ \emph {et~al.}(2006)\citenamefont {Sola},
  \citenamefont {M{\'e}vel}, \citenamefont {Elouga}, \citenamefont {Constant},
  \citenamefont {Strelkov}, \citenamefont {Poletto}, \citenamefont {Villoresi},
  \citenamefont {Benedetti}, \citenamefont {Caumes}, \citenamefont {Stagira}
  \emph {et~al.}}]{sola2006controlling}%
  \BibitemOpen
  \bibfield  {author} {\bibinfo {author} {\bibfnamefont {I.}~\bibnamefont
  {Sola}}, \bibinfo {author} {\bibfnamefont {E.}~\bibnamefont {M{\'e}vel}},
  \bibinfo {author} {\bibfnamefont {L.}~\bibnamefont {Elouga}}, \bibinfo
  {author} {\bibfnamefont {E.}~\bibnamefont {Constant}}, \bibinfo {author}
  {\bibfnamefont {V.}~\bibnamefont {Strelkov}}, \bibinfo {author}
  {\bibfnamefont {L.}~\bibnamefont {Poletto}}, \bibinfo {author} {\bibfnamefont
  {P.}~\bibnamefont {Villoresi}}, \bibinfo {author} {\bibfnamefont
  {E.}~\bibnamefont {Benedetti}}, \bibinfo {author} {\bibfnamefont {J.-P.}\
  \bibnamefont {Caumes}}, \bibinfo {author} {\bibfnamefont {S.}~\bibnamefont
  {Stagira}}, \emph {et~al.},\ }\bibfield  {title} {\bibinfo {title}
  {Controlling attosecond electron dynamics by phase-stabilized polarization
  gating},\ }\href@noop {} {\bibfield  {journal} {\bibinfo  {journal} {Nature
  Physics}\ }\textbf {\bibinfo {volume} {2}},\ \bibinfo {pages} {319} (\bibinfo
  {year} {2006})}\BibitemShut {NoStop}%
\bibitem [{\citenamefont {Walecki}\ \emph {et~al.}(1997)\citenamefont
  {Walecki}, \citenamefont {Fittinghoff}, \citenamefont {Smirl},\ and\
  \citenamefont {Trebino}}]{Walecki:97}%
  \BibitemOpen
  \bibfield  {author} {\bibinfo {author} {\bibfnamefont {W.~J.}\ \bibnamefont
  {Walecki}}, \bibinfo {author} {\bibfnamefont {D.~N.}\ \bibnamefont
  {Fittinghoff}}, \bibinfo {author} {\bibfnamefont {A.~L.}\ \bibnamefont
  {Smirl}},\ and\ \bibinfo {author} {\bibfnamefont {R.}~\bibnamefont
  {Trebino}},\ }\bibfield  {title} {\bibinfo {title} {Characterization of the
  polarization state of weak ultrashort coherent signals by dual-channel
  spectral interferometry},\ }\href {https://doi.org/10.1364/OL.22.000081}
  {\bibfield  {journal} {\bibinfo  {journal} {Opt. Lett.}\ }\textbf {\bibinfo
  {volume} {22}},\ \bibinfo {pages} {81} (\bibinfo {year} {1997})}\BibitemShut
  {NoStop}%
\bibitem [{\citenamefont {Dennis}\ \emph {et~al.}(2009)\citenamefont {Dennis},
  \citenamefont {O'Holleran},\ and\ \citenamefont
  {Padgett}}]{dennis_chapter_2009}%
  \BibitemOpen
  \bibfield  {author} {\bibinfo {author} {\bibfnamefont {M.~R.}\ \bibnamefont
  {Dennis}}, \bibinfo {author} {\bibfnamefont {K.}~\bibnamefont {O'Holleran}},\
  and\ \bibinfo {author} {\bibfnamefont {M.~J.}\ \bibnamefont {Padgett}},\
  }\bibfield  {title} {\bibinfo {title} {Chapter 5 {Singular} {Optics}:
  {Optical} {Vortices} and {Polarization} {Singularities}},\ }in\ \href
  {https://doi.org/10.1016/S0079-6638(08)00205-9} {\emph {\bibinfo {booktitle}
  {Progress in Optics}}},\ Vol.~\bibinfo {volume} {53}\ (\bibinfo  {publisher}
  {Elsevier},\ \bibinfo {year} {2009})\ pp.\ \bibinfo {pages}
  {293--363}\BibitemShut {NoStop}%
\end{thebibliography}%

\end{document}